\documentclass[prl, twocolumn, superscriptaddress]{revtex4}
\usepackage{graphicx}
\usepackage{amsmath}
\usepackage{amssymb}
\usepackage{bm}
\usepackage{color}
\usepackage{hyperref}
\usepackage{tabularx}
\usepackage{multirow}
\usepackage{braket}
\usepackage{xcolor}

\begin{document}

\title{Superconducting Orbital Magnetoelectric Effect and its Evolution across the Superconductor-Normal Metal Phase Transition}
\author{Wen-Yu He} \thanks{wenyuhe@mit.edu}
\affiliation{Department of Physics, Massachusetts Institute of Technology, Cambridge, Massachusetts 02139, USA} 
\author{K. T. Law} \thanks{phlaw@ust.hk}
\affiliation{Department of Physics, Hong Kong University of Science and Technology, Clear Water Bay, Hong Kong, China}
\date{\today}

\begin{abstract}
{Superconducting magnetoelectric effect, which is the current-induced magnetization in a superconductor, mainly focused on the spin magnetization in previous studies, but ignore the effect of the orbital magnetic moments carried by the paired Bloch electrons. In this work, we show that orbital magnetic moments in superconductors can induce large orbital magnetization in the presence of a current. We constructed a unified description for the current-induced spin and orbital magnetization across the superconductor-normal metal phase transition. We find that in a superconductor with uniform pairing, the current-induced magnetization at a given current density is the same as that in its normal metal state, while with the nonuniform superconducting pairing, the current-induced magnetization exhibits an abrupt change in magnitude near the superconductor-normal metal phase transition. Importantly, our theory predicts the orbital magnetoelectric effect in superconducting twisted bilayer graphene which has paired Bloch electrons with large orbital magnetic moments and negligible spin-orbit coupling.  We propose that the measurement of the current-induced orbital magnetoelectric effect can be used to detect the possible nonuniform pairings in twisted bilayer graphene and other newly discovered superconductors with non-trivial Berry curvatures.}
\end{abstract}

\maketitle
\emph{Introduction}. Superconducting magnetoelectric effect is the current-induced magnetization in the superconducting state of a material. In previous studies, it mainly focused on the current-induced spin magnetization which arises from the spin-orbit coupling (SOC) in noncentrosymmetric superconductors~\cite{Levitov, Edelstein1, Edelstein2, Yip, Samokhin, Fujimoto, Sigrist, Tkachov, Wenyu1}. Besides the SOC, nonzero Berry curvature can also arise due to the inversion symmetry breaking~\cite{QianNiu1} and has effect on the superconducting state~\cite{Zhiwang}. It is known that Berry curvature introduces orbital magnetic moments to the Bloch electrons~\cite{QianNiu1, QianNiu2, QianNiu3, QianNiu4, Resta1, Resta2}, and the orbital magnetic moments are the source of the current-induced orbital magnetization in a normal metal~\cite{Moore, Pesin}. In the superconducting state, the orbital magnetic moments from the paired Bloch electrons are involved in Cooper pairs, but the effect of orbital magnetic moments has never been studied in the superconducting magnetoelectric effect. It raises the problem of how the current-induced orbital magnetization would evolve in the phase transition from the normal metal state to the superconducting state.

In this work, we show that in superconductors where the paired Bloch electrons carry orbital magnetic moments, applying current can generate orbital magnetization. Importantly, we constructed a unified description for the current-induced spin and orbital magnetization that is applicable in both regions across the superconductor-normal metal phase transition. At the phase transition, the current-induced magnetization in the normal metal state is found to have smooth connection to that in the superconducting state when the pairing is uniform on the Fermi surfaces, while the magnetization exhibits an abrupt change in the nonuniform pairing case.  The abrupt change is further ascribed to the nonuniform excitation of quasiparticles controlled by the nonuniform pairing on the Fermi surfaces. 

To demonstrate the utility of our theory, we study the case of twisted bilayer graphene (TBG) and predict that the TBG exhibits current-induced orbital magnetization in the superconductivity region. The TBG has recently been observed to exhibit both superconductivity\cite{Caoyuan1, Yankowitz, Lau, Efetov, Young1, Efetov2} and orbital magnetism \cite{Efetov, David, Young2, Young3, Young4}. Importantly, the superconductivity and orbital magnetism have been observed in the same TBG sample (at different filling factors)~\cite{Efetov}, indicating that the Cooper pairs are formed by electrons carrying finite orbital magnetic moments. Here we find that the orbital magnetic moments from the paired Bloch electrons can give rise to current-induced orbital magnetization in superconducting TBG. More importantly, we point out that the measurement of current-induced orbital magnetization across the superconductor-normal metal phase transition in TBG can test whether superconducting TBG has uniform or nonuniform pairing order parameters. Besides being applicable to TBG, our theory is generally applicable to a large number of noncentrosymmetric superconductors with finite Berry curvatures.

\begin{figure}
\centering
\includegraphics[width=3.2in]{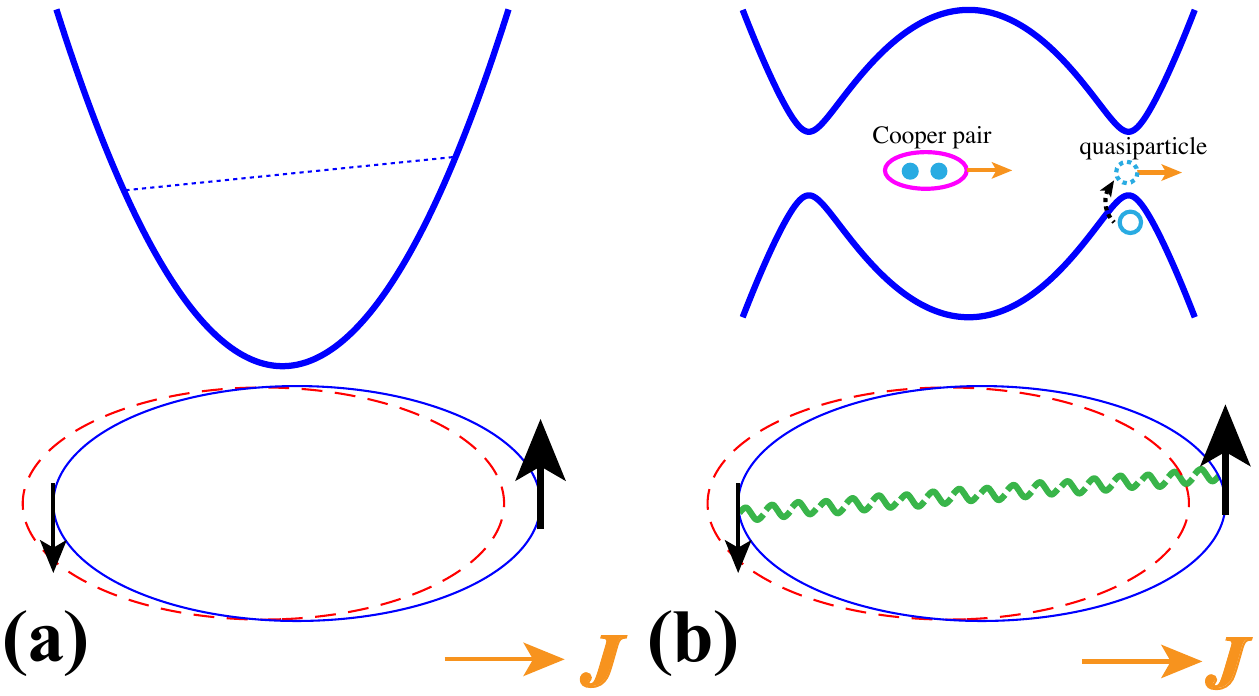}
\caption{Applying a current to a normal metal (a) and a superconductor (b). The blue and red circle correspond to the Fermi surface with and without current respectively. In a normal metal, applying a current makes more electronic states with momentum parallel to the applied current density $\bm{J}$ occupied at the Fermi energy. The occupation is schematically illustrated by the blue dot line in (a). As Bloch electrons carry magnetic moments, the redistribution at the Fermi energy gives rise to net magnetization. In a superconductor, the Bloch electrons of net momentum get paired to form Cooper pairs of net momentum. The magnetic moments carried by the paired Bloch electrons are involved in the pairing condensation and can generate a net magnetization. At finite temperature, there are quasiparticle excitations in the superconducting state, so the excited quasiparticles also contribute to the current and current-induced magnetization at $0<T<T_{\textrm{c}}$. The black arrow denotes the total magnetic moments carried by the electrons moving forward or backward, and the amplitude of the magnetic moment is schematically represented by the size of the arrow.}\label{figure1}
\end{figure}

\emph{A unified description for the magnetoelectric effect}. Both the normal and superconducting states of a material can be described by the Bogliubov-de Gennes Hamiltonian~\cite{SM}
\begin{align}
\mathcal{H}=\frac{1}{2}\sum_{\bm{k}}\begin{pmatrix}
c^\dagger_{\bm{k}} & c_{-\bm{k}}
\end{pmatrix}\begin{pmatrix}
H_0\left(\bm{k}\right) & \hat{\Delta}\left(\bm{k}\right) \\
\hat{\Delta}^\dagger\left(\bm{k}\right) & -H_0^\ast\left(-\bm{k}\right)
\end{pmatrix}\begin{pmatrix}
c_{\bm{k}} \\ c^\dagger_{-\bm{k}}
\end{pmatrix}.
\end{align}
Here $H_0\left(\bm{k}\right)$ is the normal state Hamiltonian matrix, $\hat{\Delta}\left(\bm{k}\right)$ is the pairing matrix and $c^\dagger_{\bm{k}}\left(c_{\bm{k}}\right)$ is the creation (annihilation) operator that includes multiple components for all the orbital and spin degrees of the system. In the normal state, a Bloch electronic state $\ket{\phi_{\nu, \bm{k}}}$ with energy $\xi_{\nu, \bm{k}}=\bra{\phi_{\nu, \bm{k}}}H_0\left(\bm{k}\right)\ket{\phi_{\nu, \bm{k}}}$ carries the spin magnetic moment $\bm{S}_{\nu, \bm{k}}=\bra{\phi_{\nu, \bm{k}}}\frac{1}{2}\mu_{\textrm{B}}g\bm{\sigma}\ket{\phi_{\nu, \bm{k}}}$, with $\mu_{\textrm{B}}$ being the Bohr magneton and $g$ the Lande $g$ factor. The SOC in the material is manifested by the locking of the spin magnetic moment direction with the group velocity $\bm{v}_{\nu, \bm{k}}=\partial_{\bm{k}}\xi_{\nu, \bm{k}}$. In the absence of inversion symmetry, nonzero Berry curvature $\bm{\Omega}_{\nu, \bm{k}}=i\bra{\partial_{\bm{k}}\phi_{\nu, \bm{k}}}\times\ket{\partial_{\bm{k}}\phi_{\nu, \bm{k}}}$ can arise and endow the Bloch electronic state with finite orbital magnetic moment $\bm{m}_{\nu, \bm{k}}=\frac{ie}{2\hbar}\bra{\partial_{\bm{k}}\phi_{\nu, \bm{k}}}\times\left[H_0\left(\bm{k}\right)-\xi_{\nu, \bm{k}}\right]\ket{\partial_{\bm{k}}\phi_{\nu, \bm{k}}}$~\cite{QianNiu1, QianNiu2, QianNiu3}. Therefore, the total magnetic moment of a Bloch electron is $\bm{M}_{\nu, \bm{k}}=\bm{S}_{\nu, \bm{k}}+\bm{m}_{\nu, \bm{k}}$. In the superconducting state with time reversal symmetry, as a Bloch electronic state $\ket{\phi_{\nu, \bm{k}}}$ always has a partner $\ket{\phi_{\nu, -\bm{k}}}$ with the same energy $\xi_{\nu, \bm{k}}=\xi_{\nu, -\bm{k}}$, the two Bloch electronic states can pair and give rise to the pairing order parameter $\Delta_{\nu, \bm{k}}\sim\left\langle \phi_{\nu, -\bm{k}}\phi_{\nu, \bm{k}}\right\rangle$~\cite{Sigrist1, SM}, yielding the Bogliubov quasiparticle spectrum $\epsilon_{\nu, \bm{k}}=\sqrt{\xi^2_{\nu, \bm{k}}+|\Delta_{\nu, \bm{k}}|^2}$. Therefore, in a noncentrosymmetric superconductor with finite Berry curvature, the paired electrons generally carry both spin and orbital magnetic moments even though the net magnetization is zero in the absence of a current. However, as we show below, applying a current can induce net spin and orbital magnetization.

Applying a current to a metal or superconductor can be described by introducing a $U\left(1\right)$ gauge field to the original Hamiltonian $\mathcal{H}\rightarrow\mathcal{H}+\delta\mathcal{H}$ where
\begin{align}\nonumber
\delta\mathcal{H}=&-\sum_{\bm{k}, \bm{q}}c^\dagger_{\bm{k}-\frac{1}{2}\bm{q}}c_{\bm{k}+\frac{1}{2}\bm{q}}\left[\frac{e}{\hbar}\partial_{k_i}H_0\left(\bm{k}\right)A_i\left(-\bm{q}, t\right)\right.\\
&\left.-\frac{e^2}{2\hbar^2}\partial^2_{k_ik_j}H_0\left(\bm{k}\right)A_i\left(\bm{q}, t\right)A_j\left(-\bm{q}, t\right)\right],
\end{align}
with the spatial components denoted by $i, j= x, y, z$, and $\bm{A}\left(\bm{r}, t\right)=\sum_{\bm{q}}\bm{A}\left(\bm{q}, t\right)e^{i\bm{q}\cdot\bm{r}}$ being the vector gauge potential. The perturbation $\delta\mathcal{H}$ changes the distribution of Bloch electrons at the Fermi energy, and also changes the original Fermi surfaces, as is shown in Fig. \ref{figure1} (a). In the normal metallic state, the redistribution of Bloch electrons results in an imbalance of the total magnetic moments of the occupied states, so a net magnetization arises by applying a current. In the superconducting state, the Bloch electrons on the Fermi surfaces with net momentum $\bm{q}$ are paired into Cooper pairs, which have the form $\Delta_{\nu, \bm{k}, \bm{q}}\sim\left\langle \phi_{\nu, -\bm{k}+\frac{1}{2}\bm{q}}\phi_{\nu, \bm{k}+\frac{1}{2}\bm{q}} \right\rangle$. The Cooper pairs with net momentum generate supercurrent, and also induce net magnetization that comes from the spin and orbital magnetic moments in the paired Bloch electrons. Besides the Cooper pairs, quasiparticles with net momentum are excited at finite temperature and also contribute to the net magnetization. Therefore the net magnetization in a superconductor comes from two aspects: one is the supercurrent and the other is the quasiparticle current, as is schematically illustrated in Fig. \ref{figure1} (b). Importantly, the way for a supercurrent to induce net magnetization is in analogy to that in the normal metallic state: in the superconductivity the supercurrent accumulates net spin and orbital magnetic moments in the pairing condensation, while in the normal metal the current populates net magnetic moments at the Fermi energy. 

To calculate the current-induced magnetization in both the normal and the superconducting states, we apply the linear response theory and obtain the current-induced bulk magnetization as~\cite{SM}
\begin{align}\label{magnetization}\nonumber
M_i=&-\frac{e}{\hbar}\sum_{\nu}\int_{\bm{k}}M_{\nu, \bm{k}, i}v_{\nu, \bm{k}, j}\left\{\frac{\tau}{1-i\omega\tau}\frac{df\left(\epsilon_{\nu, \bm{k}}\right)}{d\epsilon_{\nu, \bm{k}}}E_j\right.\\
&\left.+\left[\frac{df\left(\epsilon_{\nu, \bm{k}}\right)}{d\epsilon_{\nu, \bm{k}}}-\frac{df\left(\xi_{\nu, \bm{k}}\right)}{d\xi_{\nu, \bm{k}}}\right]A_j\right\}.
\end{align}
This magnetization is associated with the current density
\begin{align}\label{current}\nonumber
J_i=&-\frac{e^2}{\hbar^2}\sum_{\nu}\int_{\bm{k}}v_{\nu, \bm{k}, i}v_{\nu, \bm{k}, j}\left\{\frac{\tau}{1-i\omega\tau}\frac{df\left(\epsilon_{\nu, \bm{k}}\right)}{d\epsilon_{\nu, \bm{k}}}E_j\right.\\
&\left.+\left[\frac{df\left(\epsilon_{\nu, \bm{k}}\right)}{d\epsilon_{\nu, \bm{k}}}-\frac{df\left(\xi_{\nu, \bm{k}}\right)}{d\xi_{\nu, \bm{k}}}\right]A_j\right\}.
\end{align}
Here, $\int_{\bm{k}}\equiv\int_{\textrm{BZ}}d\bm{k}/\left(2\pi\right)^d$ with $d$ being the dimension, $\tau$ is the effective scattering time, and $f\left(\epsilon\right)$ is the Fermi Dirac distribution function. In the derivation of Eq. \ref{magnetization} and \ref{current} we applied the Coulomb gauge so the electric field takes the form $\bm{E}=i\omega\bm{A}$~\cite{Schrieffer}. The bulk magnetization derived in Eq. \ref{magnetization} involves both the spin and orbital magnetization, and it is applicable in both the normal metal and the superconductivity region. In the normal metal region with zero pairing gap $\Delta_{\nu, \bm{k}}=0$, Eq. \ref{current} is the standard Drude formula to describe the current density under an applied electric field $\bm{E}$, and the current-induced magnetization in Eq. \ref{magnetization} recovers the magnetoelectric susceptibility of a metal~\cite{Moore, Pesin}. In the superconductivity region with finite pairing gap $|\Delta_{\nu, \bm{k}}|$, the first terms in Eq. \ref{magnetization} and \ref{current} describe the net magnetization and the associated current density from the excited quasiparticles. As the quasiparticle excitations are suppressed by the pairing gap, the current and magnetization that come from the quasiparticles approach to zero at low temperature. The second terms in Eq. \ref{magnetization} and \ref{current} describe the magnetization and supercurrent density from Cooper pairs. The supercurrent in Eq. \ref{current} arises from the vector gauge field $\bm{A}$, which twists the phase of pairing condensation. The second term in Eq. \ref{magnetization} is the net magnetization of pairing condensation induced by the supercurrent. Importantly, as the temperature decreases, the supercurrent and the supercurrent induced magnetization become dominant.

For the current-induced magnetization described in Eq. \ref{magnetization} and \ref{current}, the vector fields $\bm{E}$ and $\bm{A}$ can further be replaced by the current density $\bm{J}$, which results in the following expression:
\begin{align}\label{magnetization2}
M_i=\alpha_{ik}J_k\quad\textrm{with}\quad \alpha_{ik}=\gamma_{ij}\left(\tilde{v}^{-1}\right)_{jk}.
\end{align}
Here the susceptibility tensor $\alpha_{ij}$ describes the magnetization induced by a unit current density and can be nonzero only for crystals with point group symmetry belonging to one of the 18 gyrotropic point groups~\cite{Wenyu1, Moore0}. The tensors $\gamma_{ij}$, $\tilde{v}_{jk}$ that are used to calculate $\alpha_{ij}$ are
\begin{align}\label{tensor}
\gamma_{ij}=&\frac{e}{\hbar}\sum_\nu\int_{\bm{k}}M_{\nu, \bm{k}, i}v_{\nu, \bm{k}, j}\left[\frac{df\left(\xi_{\nu, \bm{k}}\right)}{d\xi_{\nu, \bm{k}}}-\frac{1}{1-i\omega\tau}\frac{df\left(\epsilon_{\nu, \bm{k}}\right)}{d\epsilon_{\nu, \bm{k}}}\right], \\\label{tensor2}
\tilde{v}_{jk}=&\frac{e^2}{\hbar^2}\sum_\nu\int_{\bm{k}}v_{\nu, \bm{k}, j}v_{\nu, \bm{k}, k}\left[\frac{df\left(\xi_{\nu, \bm{k}}\right)}{d\xi_{\nu, \bm{k}}}-\frac{1}{1-i\omega\tau}\frac{df\left(\epsilon_{\nu, \bm{k}}\right)}{d\epsilon_{\nu, \bm{k}}}\right].
\end{align}
Importantly, at zero temperature $T=0$ K, it is found that $\gamma_{ij}\sim\frac{e}{\hbar}\sum_\nu\oint M_{\nu, \bm{k}_{\textrm{F}}, i}v_{\nu, \bm{k}_{\textrm{F}}, j}d\bm{k}_{\textrm{F}}$ and $\tilde{v}_{ij}\sim\frac{e^2}{\hbar^2}\sum_\nu\oint v_{\nu, \bm{k}_{\textrm{F}}, i}v_{\nu, \bm{k}_{\textrm{F}}, j}d\bm{k}_{\textrm{F}}$~\cite{SM}, with $\bm{k}_{\textrm{F}}$ being the wave vector on the Fermi surfaces, so the induced magnetization $\bm{M}$ at a given current density $\bm{J}$ at $T=0$ K is the same regardless of whether the system is superconducting or not. The susceptibility $\alpha_{ij}$ at $T=0$ K is only determined by the group velocity and total magnetic moments of Bloch electrons on the Fermi surfaces. It indicates that for a normal metal that has current-induced orbital magnetization, the orbital magnetization can also arise from applying supercurrent in its superconductivity region. In the temperature range $0<T<T_{\textrm{c}}$ ($T_{\textrm{c}}$ is the critical temperature for a superconductor), as the excited quasiparticles come into play a role in the current-induced magnetization, the pairing gap that controls the quasiparticle excitations becomes another factor that can affect the magnetization at given current density.

\emph{Nonuniform pairing induced abrupt change in the current-induced magnetization}. The effect of quasiparticle excitations on the current-induced magnetization is the most prominent at the temperature near the superconductor-normal metal phase transition, where the small pairing gap near $T_{\textrm{c}}$ allows a large number of quasiparticle excitations to coexist with Cooper pairs. From the first term in Eq. \ref{magnetization} and \ref{current} it is known that the current from the excited quasiparticles and the affiliated magnetization inherit from those in the normal metal state. In the case of uniform pairing gap $|\Delta_{\nu, \bm{k}_{\textrm{F}}}|=\Delta_0$, since the suppression of quasiparticle excitations is uniform on the Fermi surfaces, the quasiparticle current and the affiliated magnetization are produced in the same ratio as the current and the current-induced magnetization in the normal metal state. In contrast, when the pairing gap $\Delta_{\nu, \bm{k}_{\textrm{F}}}$ is nonuniform on Fermi surfaces, the ratio of quasiparticle current and the affiliated magnetization differs from that in the normal metal. We know from the $T=0$ K result that the ratio of the supercurrent and the supercurrent-induced magnetization is the same as the ratio in the normal metal state. Therefore, across the superconductor-normal metal phase transition, the current-induced magnetization is exactly the same in the uniform pairing case, while an abrupt change in magnitude can happen in the nonuniform pairing case. 

\begin{figure}
\centering
\includegraphics[width=3.5in]{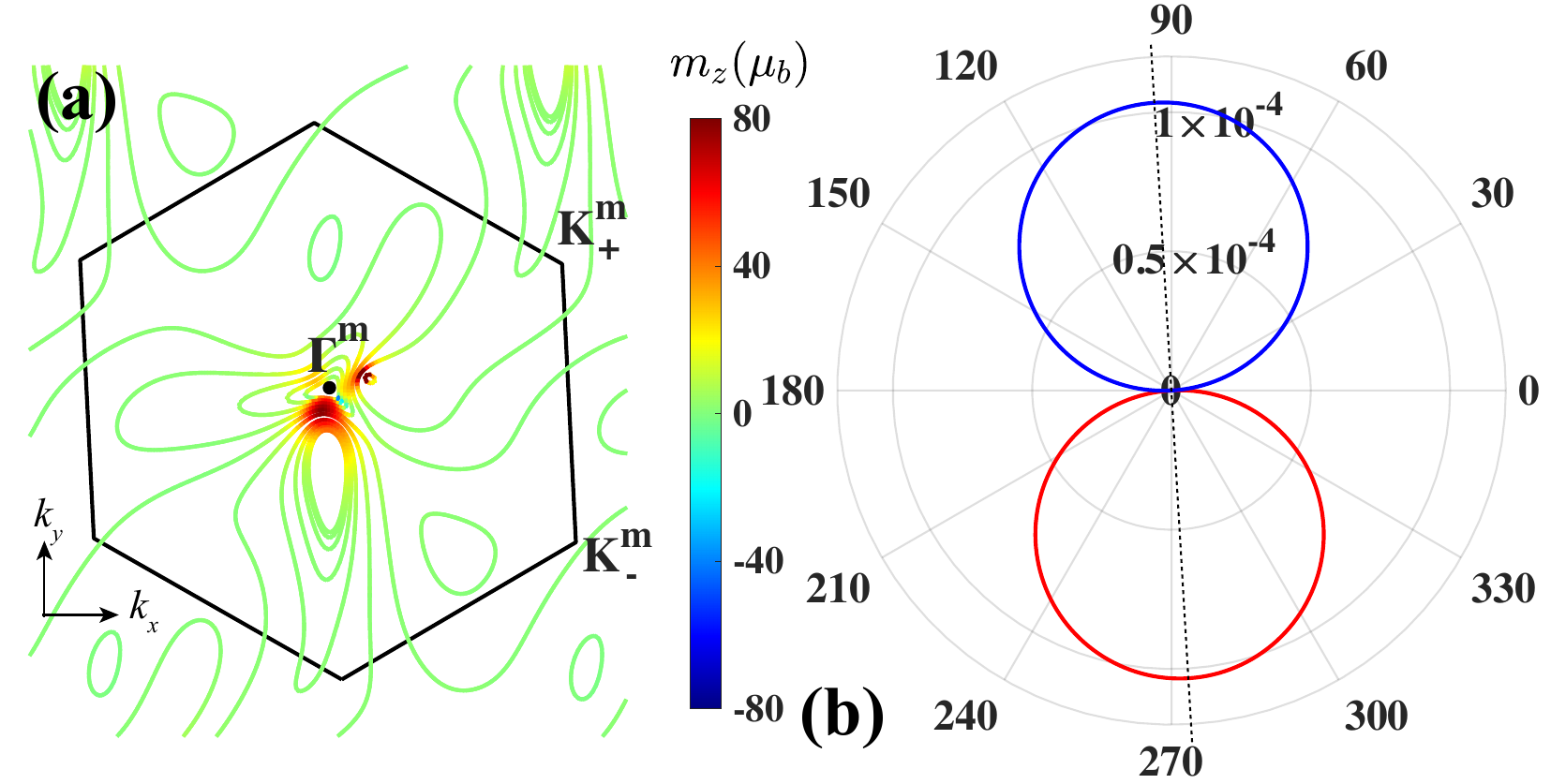}
\caption{(a) The orbital magnetic moments at different energy contours of the valence band. The tilted hexagon is the mini-Brillouin zone. Here orbital magnetic moment distribution in one valley is shown, and it can be mapped to the case in the other valley through the time reversal symmetry. (b) The out-of-plane orbital magnetization $M_z$ (in units of $\mu_{\textrm{B}}$/nm$^2$) induced by in-plane current of 1nA/nm along different directions. The optimal direction that generates the largest $M_z$ is labelled by a black dashed line. The red and blue lines mean positive and negative values of orbital magnetization $M_z$ respectively. The $x$ direction has been defined to be along the angular bisector between the two zig-zag directions of the top and bottom graphene layers.}\label{figure2}
\end{figure}

The abrupt change of current-induced magnetization at $T_{\textrm{c}}$ in the nonuniform pairing case can be deduced from the Taylor expansion of $\gamma_{ij}$, $\tilde{v}_{ij}$ in terms of $\Delta_{\nu, \bm{k}}$, which yields $\gamma_{ij}\sim\frac{e}{\hbar}\sum_\nu\oint M_{\nu, \bm{k}_{\textrm{F}}}v_{\nu, \bm{k}_{\textrm{F}}, j}|\Delta_{\nu, \bm{k}_{\textrm{F}}}|^2d\bm{k}_{\textrm{F}}$ and $\tilde{v}_{ij}\sim\frac{e^2}{\hbar^2}\sum_\nu\oint v_{\nu, \bm{k}_{\textrm{F}}, i}v_{\nu, \bm{k}_{\textrm{F}}, j}|\Delta_{\nu, \bm{k}_{\textrm{F}}}|^2d\bm{k}_{\textrm{F}}$~\cite{SM}. Above $T_{\textrm{c}}$, we know $\gamma_{ij}\sim\frac{e}{\hbar}\sum_\nu\oint M_{\nu, \bm{k}_{\textrm{F}}, i}v_{\nu, \bm{k}_{\textrm{F}}, j}d\bm{k}_{\textrm{F}}$ and $\tilde{v}_{ij}\sim\frac{e^2}{\hbar^2}\sum_\nu\oint v_{\nu, \bm{k}_{\textrm{F}}, i}v_{\nu, \bm{k}_{\textrm{F}}, j}d\bm{k}_{\textrm{F}}$~\cite{SM}, so the dependence on the paring $\Delta_{\nu, \bm{k}_{\textrm{F}}}$ arises abruptly in the susceptibility $\alpha_{ij}$ once $T<T_{\textrm{c}}$. When the pairing is nonuniform on the Fermi surfaces, the susceptibility $\alpha_{ij}$ thus differs at the two sides of $T_{\textrm{c}}$. Such abrupt change of current-induced magnetization across $T_{\textrm{c}}$ is an indicator of the nonuniform pairing in the superconducting state.

\emph{The orbital magnetoelectric effect in superconducting twisted bilayer graphene}. In previous studies, current-induced magnetization in the superconducting state only involves the spin magnetization caused by SOC~\cite{Levitov, Edelstein1, Edelstein2, Yip, Samokhin, Fujimoto, Sigrist, Tkachov, Wenyu1}. Recently, superconductivity was observed in TBG which has negligibly small SOC. Here, we predict that a large orbital magnetization can be induced by a current in superconducting TBG. More importantly, the behavior of the current-induced orbital magnetization in TBG near $T_{\textrm{c}}$ provides further information about the pairing gap in TBG.

The TBG that shows superconductivity in experiment lies on the hBN substrate, so the coupling of the bottom graphene layer with the hBN can gap out the Dirac points and generate finite Berry curvature~\cite{MacDonald, Senthil1, Senthil2, Zaletel}. Due to the mismatch between the hBN and the bottom graphene layer, the equilateral triangle Moir\'e superlattice has been observed to get deformed~\cite{Perge, Pasupathy, Yazdani}, indicating that the C$_3$ symmetry of the TBG is generally broken by the strain from the hBN substrate. The resulting TBG system has the lowest C$_1$ symmetry with finite Berry curvature, so an in-plane current can give rise to an out-of-plane orbital magnetization $M_{z}=\alpha_{zx}J_x+\alpha_{zy}J_y$ in the normal state ~\cite{Wenyu2}. Since at the $T=0$ K the current-induced magnetization in the normal state is the same as the supercurrent-induced magnetization, the current-induced orbital magnetization in TBG is always maintained in its superconducting state. To demonstrate the in-plane current-induced out-of-plane orbital magnetization in a superconducting TBG on hBN, we construct the continuum model for a TBG with a twist angle of $1.6^\circ$~\cite{Neto1, Neto2, MacDonald2}. In the TBG, there is a uniaxial strain applied along the zig-zag direction in the bottom graphene layer to have it stretched by 0.1\% (the detailed description for the strained TBG can be found in the Supplemental Material~\cite{SM}). With the massive Dirac gap of 34meV~\cite{David2, Jung}, the Bloch electrons in the valence Moir\'e flat band carry orbital magnetic moments up to 80 $\mu_{\textrm{B}}$ as shown in Fig. \ref{figure2} (a). Around 1/2 filling in the valence Moir\'e flat band, the in-plane supercurrent induced out-of-plane orbital magnetization at $T=0K$ is directly computed through Eq. \ref{tensor} and \ref{tensor2}. Assuming the in-plane supercurrent density to be 1nA/nm, the out-of-plane orbital magenetization as a function of the in-plane supercurrent direction is shown in Fig. \ref{figure2} (b). The largest orbital magnetization induced by the in-plane current along the optimal direction reaches the order of $10^{-4}\mu_{\textrm{B}}/$nm$^2$. This magnitude of orbital magnetization is comparable to the current induced spin polarization in large Rashba SOC materials under $E=10^3\sim10^4$V/m, such as Au (111) surfaces and Bi/Ag bilayers~\cite{Johansson1, Johansson2}.

\begin{figure}
\centering
\includegraphics[width=3.5in]{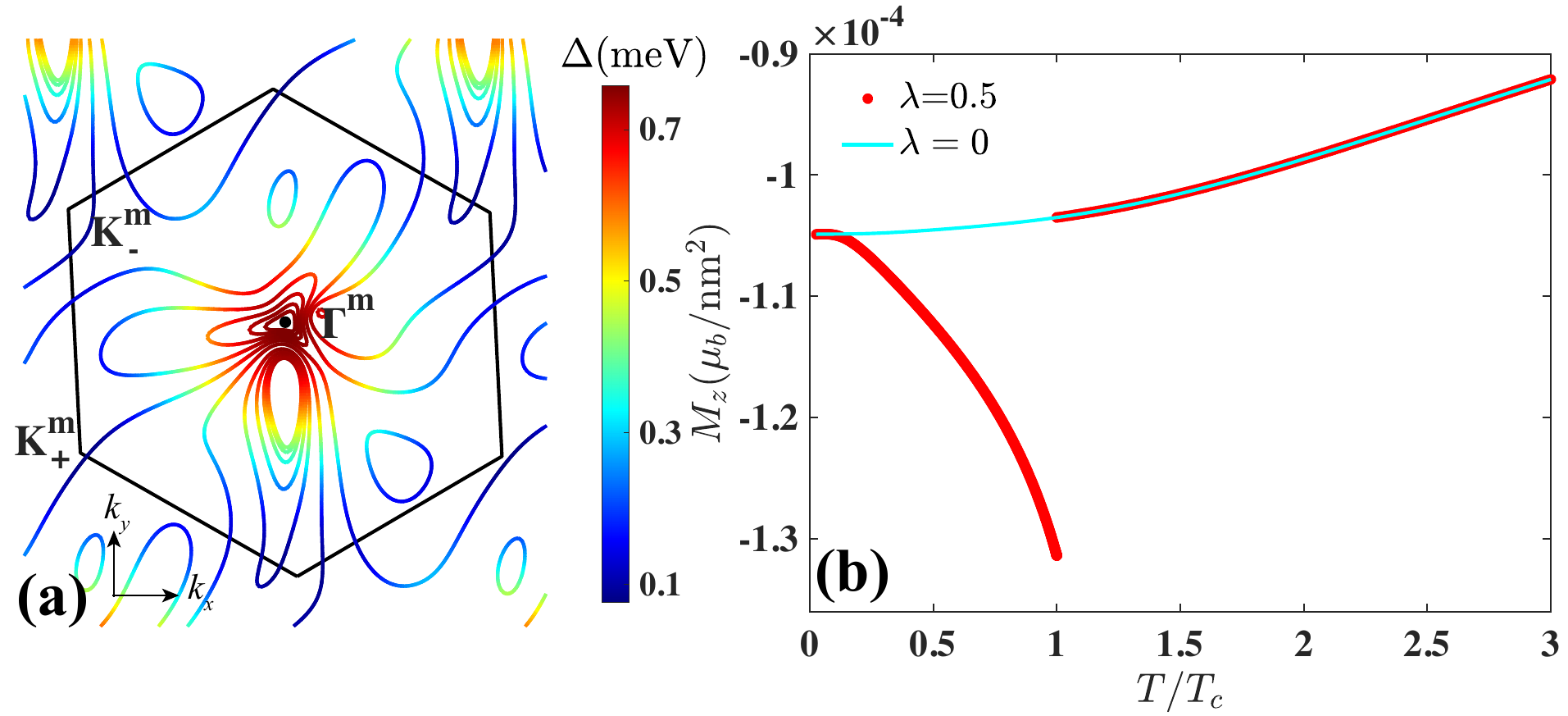}
\caption{(a) The pairing gap at different energy contours of the valence band. The pairing is plotted in one valley and that in the other valley can be obtained through time reversal symmetry. (b) The current-induced out-of-plane orbital magnetization across the superconductor-normal metal phase transition. The applied in-plane current density is 1nA/nm and the direction is along the optimal direction labelled in Fig. \ref{figure2} (b). The out-of-plane orbital magnetization exhibits an abrupt jump in nonuniform pairing case while it shows smooth transition across $T_{\textrm{c}}$ in uniform paring case.}\label{figure3}
\end{figure}

Near the superconductor-normal metal phase transition, the proportion of orbital magnetization that comes from the quasiparticle current becomes considerable, so the pairing gap that controls the quasiparticle excitations will affect the current-induced orbital magnetization near $T_{\textrm{c}}$. For the superconducting TBG on hBN, a possible nonuniform singlet pairing can be approximated as $\Delta_{\nu, \bm{k}}/\Delta_0=1+\lambda\left\{\cos\left(\bm{k}\cdot\tilde{\bm{a}}_1\right)+\cos\left(\bm{k}\cdot\tilde{\bm{a}}_2\right)+\cos\left[\bm{k}\cdot\left(\tilde{\bm{a}}_1-\tilde{\bm{a}}_2\right)\right]\right\}$~\cite{BiaoLian, DasSarma, Martin, FengchengWu, Heikkila}, with $\Delta_0=1.76k_{\textrm{B}}T_{\textrm{c}}\tanh\left(1.78\sqrt{T_{\textrm{c}}/T-1}\right)$ and $\tilde{\bm{a}}_1$, $\tilde{\bm{a}}_2$ being the primitive lattice vectors of the Moir\'e superlattice under strain~\cite{SM}. The nonuniform pairing gap at $\lambda=0.5$ is plotted in the mini-Brillouin zone shown in Fig. \ref{figure3} (a), and the orbital magnetization induced by the current along the optimal direction clearly exhibits an abrupt jump at $T_{\textrm{c}}$ as seen in Fig. \ref{figure3} (b). In contrast, in the uniform pairing case which has $\lambda=0$, the current-induced orbital magnetization has smooth connection across $T_{\textrm{c}}$. This is consistent with our analysis that an abrupt change of the current-induced magnetization at $T_{\textrm{c}}$ indicates the pairing is nonuniform on the Fermi surfaces.

\emph{Discussion}. Experimentally, the magnetization that involves orbital magnetic moments polarization in the normal state of materials has been observed through the optical Kerr effect~\cite{Fai}, nuclear magnetic resonance measurements~\cite{Itou}, and the superconducting quantum interference device (SQUID)~\cite{Young4}, and these techniques can all be used to detect the magnetization in the superconducting state~\cite{JingXia, Curro, Wernsdorfer}. In the recent direct image of orbital magnetism in TBG~\cite{Young4}, it can be deduced from the measurement that the surface magnetism at the order of $10^{-4}\mu_{\textrm{B}}/\textrm{nm}^2$ corresponds to the generated static magnetic field around 1nT, so a resolution of 0.1nT SQUID device will be able to detect the orbital magnetization induced by the current density at 1nA/nm in superconducting TBG. The current-induced orbital magnetization and the possible abrupt change at $T_{\textrm{c}}$ is proportional to the applied current density in the material, so the critical current density in the superconducting state sets the upper limit of the current-induced orbital magnetization below $T_{\textrm{c}}$. As a result, superconductors with larger pairing gap that have larger critical current density would be preferred to make the effect more measurable.

It is important to note that our theory applies to a large number of recently discovered superconductors with Berry curvatures and negligible SOC, such as a trilayer graphene Moir\'e superlattice on hBN~\cite{Fengwang, Kewang, Novoselov, Pablo_Herrero}, bilayer graphene/hBN superlattices~\cite{Taniguchi}, and twisted double bilayer graphene~\cite{Philip, Guangyu}. Moreover, our theory also applies to non-centrosymmetric superconductors with strong SOC and finite Berry curvatures such as chiral crystals, Li$_2$Pt$_3$B~\cite{Salamon}, Li$_2$Pd$_3$B~\cite{Hirata}, Mo$_3$Al$_2$C~\cite{Prozorov}, TaRh$_2$B$_2$ and NbRh$_2$B$_2$~\cite{Carnicom}, which were recently discovered to have superconductivity. In these chiral crystals the orbital magnetic moment carried by the Bloch electrons originates from the Weyl SOC~\cite{Hasan2}, so the spin and orbital magnetization are strongly mixed and the orbital magnetization effect cannot be ignored.

\emph{Acknowledgement}. The authors would like to thank K. F. Mak  for inspiring discussions. K. T. Law acknowledges the support of the Croucher Foundation and HKRGC through C6025-19G, 16310219 and 16309718.


\onecolumngrid
\clearpage
\begin{center}

{\bf Supplemental Material for ``Superconducting Orbital Magnetoelectric Effect and its Evolution across the Superconductivity Normal Metal Phase Transition''}

\end{center}

\maketitle
\setcounter{equation}{0}
\setcounter{figure}{0}
\setcounter{table}{0}
\setcounter{page}{1}
\makeatletter

\renewcommand{\theequation}{S\arabic{equation}}
\renewcommand{\thefigure}{S\arabic{figure}}
\renewcommand{\thetable}{S\arabic{table}}
\renewcommand{\bibnumfmt}[1]{[S#1]}
\renewcommand{\citenumfont}[1]{S#1}

\section{Bogliubov-de Gennes Hamiltonian and the Green's Function}
The generic Hamiltonian that can describe the normal state of a system takes the form
\begin{align}
\mathcal{H}_0=\sum_{\nu, \nu', \bm{k}}c^\dagger_{\nu, \bm{k}}H_{0, \nu\nu'}c_{\nu', \bm{k}},
\end{align}
with $c^\dagger_{\nu, \bm{k}}\left(c_{\nu, \bm{k}}\right)$ being the creation (or annihilation) operator, $H_{0, \nu\nu'}\left(\bm{k}\right)$ being the element of the Hamiltonian matrix $H_0\left(\bm{k}\right)$, and $\nu=1, 2, 3, \dots$ denoting the spin and orbital index of the system. The Bogliubov-de Gennes Hamiltonian that can describe the superconductivity state of a system is
\begin{align}
\mathcal{H}=\frac{1}{2}\sum_{\bm{k}}\begin{pmatrix}
c^\dagger_{\bm{k}} & c_{-\bm{k}}
\end{pmatrix}\begin{pmatrix}
H_0\left(\bm{k}\right) & \hat{\Delta}\left(\bm{k}\right) \\
\hat{\Delta}^\dagger\left(\bm{k}\right) & -H^\ast_0\left(-\bm{k}\right)
\end{pmatrix}\begin{pmatrix}
c_{\bm{k}} \\ c^\dagger_{-\bm{k}}
\end{pmatrix},
\end{align}
where $c_{\bm{k}}=\left[c_{1, \bm{k}}, c_{2, \bm{k}}, c_{3, \bm{k}}, \dots\right]^{\textrm{T}}$, $c^\dagger_{\bm{k}}=\left[c^\dagger_{1, \bm{k}}, c^\dagger_{2, \bm{k}}, c^\dagger_{3, \bm{k}}, \dots\right]$, and $\hat{\Delta}\left(\bm{k}\right)$ is the pairing matrix. We know that the normal state Hamiltonian $H_0\left(\bm{k}\right)$ can be diagonalized by an unitary transformation $U\left(\bm{k}\right)$ as
\begin{align}
U\left(\bm{k}\right)H_0\left(\bm{k}\right)U^\dagger\left(\bm{k}\right)=\textrm{diag}\left[\xi_{1, \bm{k}}, \xi_{2, \bm{k}}, \dots, \xi_{\nu, \bm{k}}\right],
\end{align}
with $U^\dagger\left(\bm{k}\right)U\left(\bm{k}\right)=1$ and the corresponding eigen states $\phi_{\nu, \bm{k}}=U_{\nu\nu'}\left(\bm{k}\right)c_{\nu', \bm{k}}$, $\phi^\dagger_{\nu, \bm{k}}=c^\dagger_{\nu', \bm{k}}U^\ast_{\nu'\nu}\left(\bm{k}\right)$. Applying the unitary transformation to the Bogliubov-de Gennes Hamiltonian then yields
\begin{align}
\begin{pmatrix}
U_{\bm{k}} & 0 \\
0 & U^\ast_{-\bm{k}}
\end{pmatrix}\begin{pmatrix}
H_0\left(\bm{k}\right) & \hat{\Delta}_{\bm{k}} \\
\hat{\Delta}^\dagger_{\bm{k}} & -H^\ast_0\left(-\bm{k}\right)
\end{pmatrix}\begin{pmatrix}
U^\dagger_{\bm{k}} & 0 \\
0 & U^{\textrm{T}}_{-\bm{k}}
\end{pmatrix}=\begin{pmatrix}
U_{\bm{k}}H_0\left(\bm{k}\right)U^\dagger_{\bm{k}} & U_{\bm{k}}\hat{\Delta}_{\bm{k}}U^{\textrm{T}}_{-\bm{k}} \\
U^\ast_{-\bm{k}}\hat{\Delta}^\dagger_{\bm{k}}U^\dagger_{\bm{k}} & -U^\ast_{-\bm{k}}H^\ast_0\left(-\bm{k}\right)U^{\textrm{T}}_{-\bm{k}}
\end{pmatrix}.
\end{align}
Importantly, after the unitary transformation, we obtain
\begin{align}
U_{\bm{k}}H_0\left(\bm{k}\right)U^\dagger_{\bm{k}}=\textrm{diag}\left[\xi_{1, \bm{k}}, \xi_{2, \bm{k}}, ..., \xi_{\nu, \bm{k}}\right],\quad\quad\quad -U^\ast_{-\bm{k}}H_0^\ast\left(-\bm{k}\right)U^{\textrm{T}}_{-\bm{k}}=\textrm{diag}\left[-\xi_{1, -\bm{k}}, -\xi_{2, -\bm{k}}, ..., -\xi_{\nu, -\bm{k}}\right].
\end{align}
In such eigen-band basis $\left[\phi^\dagger_{1, \bm{k}}, \phi^\dagger_{2, \bm{k}}, ..., \phi^\dagger_{\nu, \bm{k}}, \phi_{1, -\bm{k}}, \phi_{2, -\bm{k}}, ..., \phi_{\nu, -\bm{k}}\right]$, the intraband pairing $\left\langle \phi_{\nu, -\bm{k}}\phi_{\nu, \bm{k}} \right\rangle$ gives an energetically favourable pairing phase, while other interband pairing will require extra attractive interaction to overcome the energy (momentum) difference~\cite{Sigrist}. In the energetically favourable pairing phase, the pairing matrix respects $H_0\left(\bm{k}\right)\hat{\Delta}_{\bm{k}}-\hat{\Delta}_{\bm{k}}H^\ast_0\left(-\bm{k}\right)=0$~\cite{Sigrist}, and the pairing matrix is therefore diagonalized simultaneously by the unitary transformation as
\begin{align}
U_{\bm{k}}\hat{\Delta}_{\bm{k}}U^{\textrm{T}}_{-\bm{k}}=\textrm{diag}\left[\Delta_{1, \bm{k}}, \Delta_{2, \bm{k}}, ..., \Delta_{\nu, \bm{k}}\right],\quad\quad\quad U^\ast_{-\bm{k}}\hat{\Delta}_{\bm{k}}U^\dagger_{\bm{k}}=\textrm{diag}\left[\Delta^\ast_{1, \bm{k}}, \Delta^\ast_{2, \bm{k}}, ..., \Delta^\ast_{\nu, \bm{k}}\right].
\end{align}
As a result, the Bogliubov-de Gennes Hamiltonian in the eigen-band basis takes the form
\begin{align}
\begin{pmatrix}
U_{\bm{k}} & 0 \\
0 & U^\ast_{-\bm{k}}
\end{pmatrix}\begin{pmatrix}
H_0\left(\bm{k}\right) & \hat{\Delta}_{\bm{k}} \\
\hat{\Delta}^\dagger_{\bm{k}} & -H^\ast_0\left(-\bm{k}\right)
\end{pmatrix}\begin{pmatrix}
U^\dagger_{\bm{k}} & 0 \\
0 & U^{\textrm{T}}_{-\bm{k}}
\end{pmatrix}=\begin{pmatrix}
\xi_{1, \bm{k}} & 0 & 0 & 0 & \Delta_{1, \bm{k}} & 0 & 0 & 0 \\
0 & \xi_{2, \bm{k}} & 0 & 0 & 0 & \Delta_{2, \bm{k}} & 0 & 0 \\
0 & 0 & \ddots & 0 & 0 & 0 & \ddots & 0 \\
0 & 0 & 0 & \xi_{\nu, \bm{k}} & 0 & 0 & 0 & \Delta_{\nu, \bm{k}} \\
\Delta^\ast_{1, \bm{k}} & 0 & 0 & 0 & -\xi_{1, -\bm{k}} & 0 & 0 & 0 \\
0 & \Delta^\ast_{2, \bm{k}} & 0 & 0 & 0 & -\xi_{2, -\bm{k}} & 0 & 0 \\
0 & 0 & \ddots & 0 & 0 & 0 & \ddots & 0 \\
0 & 0 & 0 & \Delta^\ast_{\nu, \bm{k}} & 0 & 0 & 0 & -\xi_{\nu, -\bm{k}}
\end{pmatrix}.
\end{align}

We define the Green's function for the Bogliubov-de Gennes Hamiltonian as
\begin{align}
G^{-1}_0\left(\bm{k}, i\omega_n\right)=\begin{pmatrix}
i\omega_n-H_0\left(\bm{k}\right) & -\hat{\Delta}_{\bm{k}} \\
-\hat{\Delta}_{\bm{k}} & i\omega_n+H^\ast_0\left(-\bm{k}\right)
\end{pmatrix},\quad\textrm{and}\quad G_0\left(\bm{k}, i\omega_n\right)=\begin{pmatrix}
G_{e}\left(\bm{k}, i\omega_n\right) & F\left(\bm{k}, i\omega_n\right) \\
F^\dagger\left(\bm{k}, i\omega_n\right) & G_{h}\left(\bm{k}, i\omega_n\right)
\end{pmatrix}
\end{align}
and then we apply the unitary transformation to the Green's function:
\begin{align}
\begin{pmatrix}
U_{\bm{k}} & 0 \\
0 & U^\ast_{-\bm{k}}
\end{pmatrix}\begin{pmatrix}
G_{e}\left(\bm{k}, i\omega_n\right) & F\left(\bm{k}, i\omega_n\right) \\
F^\dagger\left(\bm{k}, i\omega_n\right) & G_{h}\left(\bm{k}, i\omega_n\right)
\end{pmatrix}\begin{pmatrix}
U^\dagger_{\bm{k}} & 0 \\
0 & U^{\textrm{T}}_{-\bm{k}}
\end{pmatrix}=\begin{pmatrix}
U_{\bm{k}}G_{e}\left(\bm{k}, i\omega_n\right)U^\dagger_{\bm{k}} & U_{\bm{k}}F\left(\bm{k}, i\omega_n\right)U^{\textrm{T}}_{-\bm{k}} \\
U^\ast_{-\bm{k}}F^\dagger\left(\bm{k}, i\omega_n\right)U^\dagger_{\bm{k}} & U^\ast_{-\bm{k}}G_{h}\left(\bm{k}, i\omega_n\right)U^{\textrm{T}}_{-\bm{k}}
\end{pmatrix},
\end{align}
which gives
\begin{align}\label{Green_e}
U_{\bm{k}}G_{e}\left(\bm{k}, i\omega_n\right)U^\dagger_{\bm{k}}&=\begin{pmatrix}
-\frac{i\omega_n+\xi_{1, \bm{k}}}{\omega_n^2+\xi^2_{1, \bm{k}}+\Delta_{1, \bm{k}}\Delta^\ast_{1, \bm{k}}} & 0 & 0 & 0 \\
0 & -\frac{i\omega_n+\xi_{2, \bm{k}}}{\omega_n^2+\xi^2_{2, \bm{k}}+\Delta_{2, \bm{k}}\Delta^\ast_{2, \bm{k}}} & 0 & 0 \\
0 & 0 & ... & 0 \\
0 & 0 & 0 & -\frac{i\omega_n+\xi_{\nu, \bm{k}}}{\omega_n^2+\xi^2_{\nu, \bm{k}}+\Delta_{\nu, \bm{k}}\Delta^\ast_{\nu, \bm{k}}}
\end{pmatrix},\\\label{Green_F}
U_{\bm{k}}F\left(\bm{k}, i\omega_n\right)U^{\textrm{T}}_{-\bm{k}}&=\begin{pmatrix}
-\frac{\Delta_{1, \bm{k}}}{\omega_n^2+\xi^2_{1, \bm{k}}+\Delta_{1, \bm{k}}\Delta^\ast_{1, \bm{k}}} & 0 & 0 & 0 \\
0 & -\frac{\Delta_{2, \bm{k}}}{\omega_n^2+\xi^2_{2, \bm{k}}+\Delta_{2, \bm{k}}\Delta^\ast_{2, \bm{k}}} & 0 & 0 \\
0 & 0 & ... & 0 \\
0 & 0 & 0 & -\frac{\Delta_{\nu, \bm{k}}}{\omega_n^2+\xi^2_{\nu, \bm{k}}+\Delta_{\nu, \bm{k}}\Delta^\ast_{\nu, \bm{k}}}
\end{pmatrix},\\\label{Green_FT}
U^\ast_{-\bm{k}}F^\dagger\left(\bm{k}, i\omega_n\right)U^\dagger_{\bm{k}}&=\begin{pmatrix}
-\frac{\Delta^\ast_{1, \bm{k}}}{\omega_n^2+\xi^2_{1, \bm{k}}+\Delta_{1, \bm{k}}\Delta^\ast_{1, \bm{k}}} & 0 & 0 & 0 \\
0 & -\frac{\Delta^\ast_{2, \bm{k}}}{\omega_n^2+\xi^2_{2, \bm{k}}+\Delta_{2, \bm{k}}\Delta^\ast_{2, \bm{k}}} & 0 & 0 \\
0 & 0 & ... & 0 \\
0 & 0 & 0 & -\frac{\Delta^\ast_{\nu, \bm{k}}}{\omega_n^2+\xi^2_{\nu, \bm{k}}+\Delta_{\nu, \bm{k}}\Delta^\ast_{\nu, \bm{k}}}
\end{pmatrix},\\\label{Green_h}
U^\ast_{-\bm{k}}G_{h}\left(\bm{k}, i\omega_n\right)U^{\textrm{T}}_{-\bm{k}}&=\begin{pmatrix}
-\frac{i\omega_n-\xi_{1, \bm{k}}}{\omega_n^2+\xi^2_{1, \bm{k}}+\Delta_{1, \bm{k}}\Delta^\ast_{1, \bm{k}}} & 0 & 0 & 0 \\
0 & -\frac{i\omega_n-\xi_{2, \bm{k}}}{\omega_n^2+\xi^2_{2, \bm{k}}+\Delta_{2, \bm{k}}\Delta^\ast_{2, \bm{k}}} & 0 & 0 \\
0 & 0 & ... & 0 \\
0 & 0 & 0 & -\frac{i\omega_n-\xi_{\nu, \bm{k}}}{\omega_n^2+\xi^2_{\nu, \bm{k}}+\Delta_{\nu, \bm{k}}\Delta^\ast_{\nu, \bm{k}}}
\end{pmatrix}.
\end{align}

\section{Linear Response Theory for the current-induced Magnetization}
Applying current introduces a perturbation
\begin{align}
\delta\mathcal{H}=&-\sum_{\bm{k}, \bm{q}}c^\dagger_{\bm{k}-\frac{1}{2}\bm{q}}c_{\bm{k}+\frac{1}{2}\bm{q}}\left[\frac{e}{\hbar}\partial_{k_i}H_0\left(\bm{k}\right)A_i\left(-\bm{q}, t\right)-\frac{e^2}{2\hbar^2}\partial^2_{k_ik_j}H_0\left(\bm{k}\right)A_i\left(\bm{q}, t\right)A_j\left(-\bm{q}, t\right)\right]
\end{align}
to the Bogliubov-de Gennes Hamiltonian, so the free-energy density of the system gets changed. The change of free-energy density can be obtained through expanding the electromagnetic gauge fields, and it can be expressed as
\begin{align}\nonumber\label{Free_energy_density}
\Delta F=&\frac{1}{2}\sum_{\bm{q}, m}A_i\left(-\bm{q}, -i\omega_m\right)\Pi_{ij}\left(\bm{q}, i\omega_m\right)A_j\left(\bm{q}, i\omega_m\right)\\
&+\frac{1}{2}\sum_{\bm{q}, m} B_i\left(-\bm{q}, -i\omega_m\right)T^{\textrm{s}}_{ij}\left(\bm{q}, i\omega_m\right)A_j\left(\bm{q}, i\omega_m\right)+\frac{1}{2}\sum_{\bm{q}, m}A_i\left(-\bm{q}, -i\omega_m\right)\tilde{T}_{ij}^{\textrm{s}}\left(\bm{q}, i\omega_m\right)B_j\left(\bm{q}, i\omega_m\right).
\end{align}
Here, the terms $T^{\textrm{s}}_{ij}\left(\bm{q}, i\omega_m\right)$ and $\tilde{T}^{\textrm{s}}_{ij}\left(\bm{q}, i\omega_m\right)$, where
\begin{align}\nonumber
T_{ij}^{\textrm{s}}\left(\bm{q}, i\omega_m\right)=&\frac{1}{\beta V}\sum_{\bm{k}, n}\textrm{tr}\left[G_{e}\left(\bm{k}-\frac{1}{2}\bm{q}, i\omega_n\right)\frac{1}{2}\mu_{\textrm{B}}g\sigma_iG_{e}\left(\bm{k}+\frac{1}{2}\bm{q}, i\omega_n+i\omega_m\right)\frac{e}{\hbar}\frac{\partial H_0\left(\bm{k}\right)}{\partial k_j}\right.\\
&\left.-F\left(\bm{k}-\frac{1}{2}\bm{q}, i\omega_n\right)\frac{1}{2}\mu_{\textrm{B}}g\sigma_i^\ast F^\dagger\left(\bm{k}+\frac{1}{2}\bm{q}, i\omega_n+i\omega_m\right)\frac{e}{\hbar}\frac{\partial H_0\left(\bm{k}\right)}{\partial k_j}\right],\\\nonumber
\tilde{T}_{ij}^{\textrm{s}}\left(\bm{q}, i\omega_m\right)=&\frac{1}{\beta V}\sum_{\bm{k}, n}\textrm{tr}\left[G_{e}\left(\bm{k}-\frac{1}{2}\bm{q}, i\omega_n\right)\frac{e}{\hbar}\frac{\partial H_0\left(\bm{k}\right)}{\partial k_i}G_{e}\left(\bm{k}+\frac{1}{2}\bm{q}, i\omega_n+i\omega_m\right)\frac{1}{2}\mu_{\textrm{B}}g\sigma_j\right.\\
&\left.-F\left(\bm{k}-\frac{1}{2}\bm{q}, i\omega_n\right)\frac{e}{\hbar}\frac{\partial H^\ast_0\left(-\bm{k}\right)}{-\partial k_i}F^\dagger_0\left(\bm{k}+\frac{1}{2}\bm{q}, i\omega_n+i\omega_m\right)\frac{1}{2}\mu_{\textrm{B}}g\sigma_j\right].
\end{align}
give the spin magnetoelectric susceptibility in both the superconductivity and normal metal regions, while the polarization tensor $\Pi_{ij}\left(\bm{q}, i\omega_m\right)$, where
\begin{align}\nonumber\label{Polarization_tensor}
\Pi_{ij}\left(\bm{q}, i\omega_m\right)=&\frac{1}{\beta V}\sum_{\bm{k}, n}\textrm{tr}\left[G_{e}\left(\bm{k}, i\omega_n\right)\frac{e^2}{\hbar^2}\frac{\partial^2H_0\left(\bm{k}\right)}{\partial k_i\partial k_j}+G_{e}\left(\bm{k}-\frac{1}{2}\bm{q}, i\omega_n\right)\frac{e}{\hbar}\frac{\partial H_0\left(\bm{k}\right)}{\partial k_i}G_{e}\left(\bm{k}+\frac{1}{2}\bm{q}, i\omega_n+i\omega_m\right)\frac{e}{\hbar}\frac{\partial H_0\left(\bm{k}\right)}{\partial k_j}\right.\\
&\left.+F\left(\bm{k}-\frac{1}{2}\bm{q}, i\omega_n\right)\frac{e}{\hbar}\frac{\partial H^\ast_0\left(-\bm{k}\right)}{\partial k_i}F^\dagger\left(\bm{k}+\frac{1}{2}\bm{q}, i\omega_n+i\omega_m\right)\frac{e}{\hbar}\frac{\partial H_0\left(\bm{k}\right)}{\partial k_j}\right],
\end{align}
is responsible for the current and current-induced orbital magnetization in both the superconductivity and normal metal regions. 

\subsection{Polarization Tensor, Current, and the Current-induced Orbital Magnetization}
Substituting Eq. \ref{Green_e}, \ref{Green_F}, \ref{Green_FT}, \ref{Green_h} into Eq. \ref{Polarization_tensor}, we can obtain the polarization tensor $\Pi_{ij}\left(\bm{q}, i\omega_m\right)$, where
\begin{align}\nonumber
\Pi_{ij}\left(\bm{q}, i\omega_m\right)=\Pi_{ij}^{\textrm{sta}}\left(\bm{q}\right)+\Pi_{ij}^{\textrm{dyn}}\left(\bm{q}, i\omega_m\right),
\end{align}
with the static part $\Pi^{\textrm{sta}}_{ij}\left(\bm{q}\right)$
\begin{align}\nonumber
\Pi^{\textrm{sta}}_{ij}\left(\bm{q}\right)=&\frac{1}{2V}\frac{e^2}{\hbar^2}\sum_{\bm{k}, \nu}\left[\frac{\epsilon_{\nu, \bm{k}}+\xi_{\nu, \bm{k}}}{\epsilon_{\nu, \bm{k}}}f\left(\epsilon_{\nu, \bm{k}}\right)+\frac{\epsilon_{\nu, \bm{k}}-\xi_{\nu, \bm{k}}}{\epsilon_{\nu, \bm{k}}}f\left(-\epsilon_{\nu, \bm{k}}\right)\right]\bra{\phi_{\nu, \bm{k}}}\partial^2_{k_ik_j}H_0\left(\bm{k}\right)\ket{\phi_{\nu, \bm{k}}}\\\nonumber
&-\frac{1}{2V}\frac{e^2}{\hbar^2}\sum_{\bm{k}, \nu, \nu'}\left[\left(1-\frac{\xi_{\nu, \bm{k}-\frac{1}{2}\bm{q}}\xi_{\nu', \bm{k}+\frac{1}{2}\bm{q}}+\Delta_{\nu, \bm{k}-\frac{1}{2}\bm{q}}\Delta^\ast_{\nu', \bm{k}+\frac{1}{2}\bm{q}}}{\epsilon_{\nu', \bm{k}+\frac{1}{2}\bm{q}}\epsilon_{\nu, \bm{k}-\frac{1}{2}\bm{q}}}\right)\frac{1-f\left(\epsilon_{\nu', \bm{k}+\frac{1}{2}\bm{q}}\right)-f\left(\epsilon_{\nu, \bm{k}-\frac{1}{2}\bm{q}}\right)}{\epsilon_{\nu', \bm{k}+\frac{1}{2}\bm{q}}+\epsilon_{\nu, \bm{k}-\frac{1}{2}\bm{q}}}\right.\\\nonumber
&\left.-\left(1+\frac{\xi_{\nu, \bm{k}-\frac{1}{2}\bm{q}}\xi_{\nu', \bm{k}+\frac{1}{2}\bm{q}}+\Delta_{\nu, \bm{k}-\frac{1}{2}\bm{q}}\Delta^\ast_{\bm{k}+\frac{1}{2}\bm{q}}}{\epsilon_{\nu', \bm{k}+\frac{1}{2}\bm{q}}\epsilon_{\nu, \bm{k}-\frac{1}{2}\bm{q}}}\right)\frac{f\left(\epsilon_{\nu', \bm{k}+\frac{1}{2}\bm{q}}\right)-f\left(\epsilon_{\nu, \bm{k}-\frac{1}{2}\bm{q}}\right)}{\epsilon_{\nu', \bm{k}+\frac{1}{2}\bm{q}}-\epsilon_{\nu, \bm{k}-\frac{1}{2}\bm{q}}}\right]\\
&\bra{\phi_{\nu, \bm{k}-\frac{1}{2}\bm{q}}}\partial_{k_i}H_0\left(\bm{k}\right)\ket{\phi_{\nu', \bm{k}+\frac{1}{2}\bm{q}}}\bra{\phi_{\nu', \bm{k}+\frac{1}{2}\bm{q}}}\partial_{k_j}H_0\left(\bm{k}\right)\ket{\phi_{\nu, \bm{k}-\frac{1}{2}\bm{q}}},
\end{align}
and the dynamic part $\Pi^{\textrm{dyn}}_{ij}\left(\bm{q}, i\omega_m\right)$, where
\begin{align}\nonumber
\Pi_{ij}^{\textrm{dyn}}\left(\bm{q}, i\omega_m\right)=&\frac{1}{4V}\sum_{\bm{k}, \nu, \nu'}\left\{\left(1-\frac{\xi_{\nu, \bm{k}-\frac{1}{2}\bm{q}}\xi_{\nu', \bm{k}+\frac{1}{2}\bm{q}}+\Delta_{\nu, \bm{k}-\frac{1}{2}\bm{q}}\Delta^\ast_{\nu', \bm{k}+\frac{1}{2}\bm{q}}}{\epsilon_{\nu', \bm{k}+\frac{1}{2}\bm{q}}\epsilon_{\nu, \bm{k}-\frac{1}{2}\bm{q}}}\right)\left[\frac{i\omega_m}{i\omega_m+\epsilon_{\nu', \bm{k}+\frac{1}{2}\bm{q}}+\epsilon_{\nu, \bm{k}-\frac{1}{2}\bm{q}}}+\frac{i\omega_m}{i\omega_m-\left(\epsilon_{\nu', \bm{k}+\frac{1}{2}\bm{q}}+\epsilon_{\nu, \bm{k}-\frac{1}{2}\bm{q}}\right)}\right]\right.\\\nonumber
&\left.\frac{1-f\left(\epsilon_{\nu', \bm{k}+\frac{1}{2}\bm{q}}\right)-f\left(\epsilon_{\nu, \bm{k}-\frac{1}{2}\bm{q}}\right)}{\epsilon_{\nu', \bm{k}+\frac{1}{2}\bm{q}}+\epsilon_{\nu, \bm{k}-\frac{1}{2}\bm{q}}}\right.\\\nonumber
&\left.-\left(1+\frac{\xi_{\nu, \bm{k}-\frac{1}{2}\bm{q}}\xi_{\nu', \bm{k}+\frac{1}{2}\bm{q}}+\Delta_{\nu, \bm{k}-\frac{1}{2}\bm{q}}\Delta^\ast_{\nu', \bm{k}+\frac{1}{2}\bm{q}}}{\epsilon_{\nu', \bm{k}+\frac{1}{2}\bm{q}}\epsilon_{\nu, \bm{k}-\frac{1}{2}\bm{q}}}\right)\left[\frac{i\omega_m}{i\omega_m-\left(\epsilon_{\nu', \bm{k}+\frac{1}{2}\bm{q}}-\epsilon_{\nu, \bm{k}-\frac{1}{2}\bm{q}}\right)}+\frac{i\omega_m}{i\omega_m+\epsilon_{\nu', \bm{k}+\frac{1}{2}\bm{q}}-\epsilon_{\nu, \bm{k}-\frac{1}{2}\bm{q}}}\right]\right.\\
&\left.\frac{f\left(\epsilon_{\nu', \bm{k}+\frac{1}{2}\bm{q}}\right)-f\left(\epsilon_{\nu, \bm{k}-\frac{1}{2}\bm{q}}\right)}{\epsilon_{\nu', \bm{k}+\frac{1}{2}\bm{q}}-\epsilon_{\nu, \bm{k}-\frac{1}{2}\bm{q}}}\right\}\bra{\phi_{\nu, \bm{k}-\frac{1}{2}\bm{q}}}\frac{e}{\hbar}\partial_{k_i}H_0\left(\bm{k}\right)\ket{\phi_{\nu', \bm{k}+\frac{1}{2}\bm{q}}}\bra{\phi_{\nu', \bm{k}+\frac{1}{2}\bm{q}}}\frac{e}{\hbar}\partial_{k_j}H_0\left(\bm{k}\right)\ket{\phi_{\nu, \bm{k}-\frac{1}{2}\bm{q}}}.
\end{align}
The static and dynamic part of the polarization tensor can then be expanded in terms of $\bm{q}$ to the linear order as
\begin{align}
\Pi_{ij}^{\textrm{sta}}\left(\bm{q}\right)=&\Pi_{ij}^{\textrm{sta}, \left(0\right)}+Q^{\textrm{sta}}_{ijl}q_l+\mathcal{O}\left(\bm{q}^2\right),\quad\Pi_{ij}^{\textrm{dyn}}\left(\bm{q}, i\omega_m\right)=\Pi_{ij}^{\textrm{dyn}, \left(0\right)}+Q^{\textrm{dyn}}_{ijl}q_l+\mathcal{O}\left(\bm{q}^2\right),
\end{align}
with
\begin{align}
Q^{\textrm{sta}}_{ijl}=\lim_{\bm{q}\rightarrow0}\frac{\Pi^{\textrm{sta}}_{ij}\left(\bm{q}\right)-\Pi^{\textrm{sta}}_{ij}\left(0\right)}{q_l},\quad Q^{\textrm{dyn}}_{ijl}=\lim_{\bm{q}\rightarrow0}\frac{\Pi^{\textrm{dyn}}_{ij}\left(\bm{q}\right)-\Pi^{\textrm{dyn}}_{ij}\left(0\right)}{q_l}.
\end{align}
The intra-band contribution to $\Pi_{ij}^{\textrm{sta}, \left(0\right)}$ is
\begin{align}\nonumber
\Pi^{\textrm{sta}, \left(0\right), \textrm{intra}}_{ij}=&\frac{1}{2V}\frac{e^2}{\hbar^2}\sum_{\bm{k}, \nu}\bra{\phi_{\nu, \bm{k}}}\partial^2_{k_ik_j}H_0\left(\bm{k}\right)\left[\frac{\epsilon_{\nu, \bm{k}}+\xi_{\nu, \bm{k}}}{\epsilon_{\nu, \bm{k}}}f\left(\epsilon_{\nu, \bm{k}}\right)+\frac{\epsilon_{\nu, \bm{k}}-\xi_{\nu, \bm{k}}}{\epsilon_{\nu, \bm{k}}}f\left(-\epsilon_{\nu, \bm{k}}\right)\right]\\\nonumber
&+\frac{1}{V}\frac{e^2}{\hbar^2}\sum_{\bm{k}, \nu}\bra{\phi_{\nu, \bm{k}}}\partial_{k_i}H_0\left(\bm{k}\right)\ket{\phi_{\nu, \bm{k}}}\bra{\phi_{\nu, \bm{k}}}\partial_{k_j}H_0\left(\bm{k}\right)\ket{\phi_{\nu, \bm{k}}}\frac{df\left(\epsilon_{\nu, \bm{k}}\right)}{d\epsilon_{\nu, \bm{k}}}\\\nonumber
&\approx\frac{1}{V}\frac{e^2}{\hbar^2}\sum_{\bm{k}, \nu}\left[\bra{\phi_{\nu, \bm{k}}}\partial^2_{k_ik_j}H_0\left(\bm{k}\right)\ket{\phi_{\nu, \bm{k}}}f\left(\xi_{\nu, \bm{k}}\right)+\bra{\phi_{\nu, \bm{k}}}\partial_{k_i}H_0\left(\bm{k}\right)\ket{\phi_{\nu, \bm{k}}}\bra{\phi_{\nu, \bm{k}}}\partial_{k_j}H_0\left(\bm{k}\right)\ket{\phi_{\nu, \bm{k}}}\frac{df\left(\epsilon_{\nu, \bm{k}}\right)}{d\epsilon_{\nu, \bm{k}}}\right].
\end{align}
For the inter-band contribution, we assume that the pairing gap $|\Delta_{\nu, \bm{k}}|$ is much smaller than the inversion symmetry breaking-induced band splitting, so the inter-band term can be approximated by that used in the normal metal phase with $\Delta_{\nu, \bm{k}}=0$:
\begin{align}\nonumber\label{approximation}
\Pi^{\textrm{sta}, \left(0\right), \textrm{inter}}_{ij}=&-\frac{1}{2V}\frac{e^2}{\hbar^2}\sum_{\bm{k}, \nu\neq\nu'}\left[\left(1-\frac{\xi_{\nu, \bm{k}}\xi_{\nu', \bm{k}}+\Delta_{\nu, \bm{k}}\Delta^\ast_{\nu', \bm{k}}}{\epsilon_{\nu', \bm{k}}\epsilon_{\nu, \bm{k}}}\right)\frac{1-f\left(\epsilon_{\nu', \bm{k}}\right)-f\left(\epsilon_{\nu, \bm{k}}\right)}{\epsilon_{\nu', \bm{k}}+\epsilon_{\nu, \bm{k}}}\right.\\\nonumber
&\left.-\left(1+\frac{\xi_{\nu, \bm{k}}\xi_{\nu', \bm{k}}+\Delta_{\nu, \bm{k}}\Delta^\ast_{\bm{k}}}{\epsilon_{\nu', \bm{k}}\epsilon_{\nu, \bm{k}}}\right)\frac{f\left(\epsilon_{\nu', \bm{k}}\right)-f\left(\epsilon_{\nu, \bm{k}}\right)}{\epsilon_{\nu', \bm{k}}-\epsilon_{\nu, \bm{k}}}\right]\bra{\phi_{\nu, \bm{k}}}\partial_{k_i}H_0\left(\bm{k}\right)\ket{\phi_{\nu', \bm{k}}}\bra{\phi_{\nu', \bm{k}}}\partial_{k_j}H_0\left(\bm{k}\right)\ket{\phi_{\nu, \bm{k}}}\\
\approx&\frac{1}{V}\frac{e^2}{\hbar^2}\sum_{\bm{k}, \nu\neq\nu'}\frac{f\left(\xi_{\nu', \bm{k}}\right)-f\left(\xi_{\nu, \bm{k}}\right)}{\xi_{\nu', \bm{k}}-\xi_{\nu, \bm{k}}}\bra{\phi_{\nu, \bm{k}}}\partial_{k_i}H_0\left(\bm{k}\right)\ket{\phi_{\nu', \bm{k}}}\bra{\phi_{\nu', \bm{k}}}\partial_{k_j}H_0\left(\bm{k}\right)\ket{\phi_{\nu, \bm{k}}}.
\end{align}
We also know from the gauge invariance that
\begin{align}\nonumber
&\frac{1}{V}\frac{e^2}{\hbar^2}\sum_{\bm{k}, \nu\neq\nu'}\frac{f\left(\xi_{\nu', \bm{k}}\right)-f\left(\xi_{\nu, \bm{k}}\right)}{\xi_{\nu', \bm{k}}-\xi_{\nu, \bm{k}}}\bra{\phi_{\nu, \bm{k}}}\partial_{k_i}H_0\left(\bm{k}\right)\ket{\phi_{\nu', \bm{k}}}\bra{\phi_{\nu', \bm{k}}}\partial_{k_j}H_0\left(\bm{k}\right)\ket{\phi_{\nu, \bm{k}}}\\
=&-\frac{1}{V}\frac{e^2}{\hbar^2}\sum_{\bm{k}, \nu}\left[\bra{\phi_{\nu, \bm{k}}}\partial_{k_i}H_0\left(\bm{k}\right)\ket{\phi_{\nu, \bm{k}}}\bra{\phi_{\nu, \bm{k}}}\partial_{k_j}H_0\left(\bm{k}\right)\ket{\phi_{\nu, \bm{k}}}\frac{df\left(\xi_{\nu, \bm{k}}\right)}{d\xi_{\nu, \bm{k}}}+\bra{\phi_{\nu, \bm{k}}}\partial^2_{k_ik_j}H_0\left(\bm{k}\right)\ket{\phi_{\nu, \bm{k}}}f\left(\xi_{\nu, \bm{k}}\right)\right],
\end{align}
so eventually we obtain the $\Pi_{ij}^{\textrm{sta}, \left(0\right)}$, where
\begin{align}\nonumber
\Pi_{ij}^{\textrm{sta}, \left(0\right)}=&\Pi^{\textrm{sta}, \left(0\right), \textrm{intra}}_{ij}+\Pi^{\textrm{sta}, \left(0\right), \textrm{inter}}_{ij}\\\nonumber
=&\frac{1}{V}\frac{e^2}{\hbar^2}\sum_{\nu, \bm{k}}\bra{\phi_{\nu, \bm{k}}}\partial_{k_i}H_0\left(\bm{k}\right)\ket{\phi_{\nu, \bm{k}}}\bra{\phi_{\nu, \bm{k}}}\partial_{k_j}H_0\left(\bm{k}\right)\ket{\phi_{\nu, \bm{k}}}\left[\frac{df\left(\epsilon_{\nu, \bm{k}}\right)}{d\epsilon_{\nu, \bm{k}}}-\frac{df\left(\xi_{\nu, \bm{k}}\right)}{d\xi_{\nu, \bm{k}}}\right]\\
=&\frac{e^2}{\hbar^2}\frac{1}{\left(2\pi\right)^d}\int_{\textrm{BZ}}d\bm{k}\sum_\nu v_{\nu, \bm{k}, i}v_{\nu, \bm{k}, j}\left[\frac{df\left(\epsilon_{\nu, \bm{k}}\right)}{d\epsilon_{\nu, \bm{k}}}-\frac{df\left(\xi_{\nu, \bm{k}}\right)}{d\xi_{\nu, \bm{k}}}\right],
\end{align}
with the group velocity $\bm{v}_{\nu, \bm{k}}=\nabla_{\bm{k}}\xi_{\nu, \bm{k}}=\bra{\phi_{\nu, \bm{k}}}\nabla_{\bm{k}}H_0\left(\bm{k}\right)\ket{\phi_{\nu, \bm{k}}}$. For the term $\Pi_{ij}^{\textrm{dyn}, \left(0\right)}$ that is $i\omega_m$ dependent, we consider the dominant intra-band contribution and take the analytic continuation $i\omega_m\rightarrow\hbar\omega+i\hbar\tau^{-1}$ to get
\begin{align}\label{dynamic_polarization}
\Pi_{ij}^{\textrm{dyn}, \left(0\right)}=\frac{e^2}{\hbar^2}\frac{i\omega\tau}{1-i\omega\tau}\sum_\nu\int_{\textrm{BZ}}v_{\nu, \bm{k}, i}v_{\nu, \bm{k}, j}\frac{df\left(\epsilon_{\nu, \bm{k}}\right)}{d\epsilon_{\nu, \bm{k}}}\frac{d\bm{k}}{\left(2\pi\right)^d}.
\end{align}

The intra-band contribution to $Q^{\textrm{sta}}_{ijl}$~\cite{Moore, JingMa} is
\begin{align}
Q^{\textrm{sta}, \textrm{intra}}_{ijl}=&\frac{i}{V}\frac{e}{\hbar}\sum_{\nu, \bm{k}}\frac{df\left(\epsilon_{\nu, \bm{k}}\right)}{d\epsilon_{\nu, \bm{k}}}\left(\partial_{k_i}\xi_{\nu, \bm{k}}\epsilon_{dlj}m_{\nu, \bm{k}, d}-\partial_{k_j}\xi_{\nu, \bm{k}}\epsilon_{dli}m_{\nu, \bm{k}, d}\right),
\end{align}
with $\bm{m}_{\nu, \bm{k}}=\frac{ie}{2\hbar}\bra{\partial_{\bm{k}}\phi_{\nu, \bm{k}}}\times\left[H_0\left(\bm{k}\right)-\xi_{\nu, \bm{k}}\right]\ket{\partial_{\bm{k}}\phi_{\nu, \bm{k}}}$ being the orbital magnetic moment. For the inter-band contribution, given that the pairing gap $|\Delta_{\nu, \bm{k}}|$ is much smaller than the band splitting, it can also be approximated by that used in the normal metal phase in a way that is similar to Eq. \ref{approximation}. We know that the summation of the intra-band and inter-band static terms in the normal metal phase gives $Q_{ijl}^{\textrm{sta}}=\frac{e^2}{\hbar^2}\int_{\textrm{BZ}} \frac{d\bm{k}}{\left(2\pi\right)^d}\sum_\nu2f\left(\epsilon_{\nu, \bm{k}}\right)\nabla_{\bm{k}}\cdot\left(\epsilon_{\nu, \bm{k}}\bm{\Omega}_{\nu, \bm{k}}\right)$, which is confirmed to be zero~\cite{Moore, JingMa}. As a result, the inter-band contribution to $Q^{\textrm{sta}}_{ijl}$ is
\begin{align}
Q^{\textrm{sta}, \textrm{inter}}_{ijl}=&-\frac{i}{V}\frac{e}{\hbar}\sum_{\nu, \bm{k}}\frac{df\left(\xi_{\nu, \bm{k}}\right)}{d\xi_{\nu, \bm{k}}}\left(\partial_{k_i}\xi_{\nu, \bm{k}}\epsilon_{dlj}m_{\nu, \bm{k}, d}-\partial_{k_j}\xi_{\nu, \bm{k}}\epsilon_{dli}m_{\nu, \bm{k}, d}\right).
\end{align}
For the dynamic part $Q_{ijl}^{\textrm{dyn}}$, we consider the dominant intra-band term and use the analytical continuation $i\omega_m\rightarrow\hbar\omega+i\hbar\tau^{-1}$ as well, and then we can obtain
\begin{align}
Q_{ijl}^{\textrm{dyn}}=&-\frac{1}{V}\frac{e}{\hbar}\frac{\omega\tau}{1-i\omega\tau}\sum_{\bm{k}, \nu}\frac{df\left(\epsilon_{\nu, \bm{k}}\right)}{d\epsilon_{\nu, \bm{k}}}\left(\partial_{k_i}\xi_{\nu, \bm{k}}\epsilon_{dlj}m_{\nu, \bm{k}, d}-\partial_{k_j}\xi_{\nu, \bm{k}}\epsilon_{dli}m_{\nu, \bm{k}, d}\right).
\end{align}

Eventually, the polarization tensor $\Pi_{ij}\left(\bm{q}, \omega\right)$ can be approximated:
\begin{align}\nonumber
\Pi_{ij}\left(\bm{q}, \omega\right)\approx&\Pi_{ij}^{\textrm{sta}, \left(0\right)}+Q^{\textrm{sta}}_{ijl}q_l+\Pi_{ij}^{\textrm{dyn}, \left(0\right)}+Q^{\textrm{dyn}}_{ijl}q_l\\\nonumber
=&\frac{e^2}{\hbar^2}\int_{\textrm{BZ}}\frac{d\bm{k}}{\left(2\pi\right)^d}\sum_\nu v_{\nu, \bm{k}, i}v_{\nu, \bm{k}, j}\left[\frac{df\left(\epsilon_{\nu, \bm{k}}\right)}{d\epsilon_{\nu, \bm{k}}}-\frac{df\left(\xi_{\nu, \bm{k}}\right)}{d\xi_{\nu, \bm{k}}}\right]+\frac{e^2}{\hbar^2}\frac{i\omega\tau}{1-i\omega\tau}\int_{\textrm{BZ}}\frac{d\bm{k}}{\left(2\pi\right)^d}\sum_\nu v_{\nu, \bm{k}, i}v_{\nu, \bm{k}, j}\frac{df\left(\epsilon_{\nu, \bm{k}}\right)}{d\epsilon_{\nu, \bm{k}}}\\\nonumber
&+\frac{i}{V}\frac{e}{\hbar}\sum_{\nu, \bm{k}}\left(v_{\nu, \bm{k}, i}\epsilon_{dlj}m_{\nu, \bm{k}, d}-v_{\nu, \bm{k}, j}\epsilon_{dli}m_{\nu, \bm{k}, d}\right)\left[\frac{df\left(\epsilon_{\nu, \bm{k}}\right)}{d\epsilon_{\nu, \bm{k}}}-\frac{df\left(\xi_{\nu, \bm{k}}\right)}{d\xi_{\nu, \bm{k}}}\right]\\
&-\frac{1}{V}\frac{e}{\hbar}\frac{\omega\tau}{1-i\omega\tau}\sum_{\nu, \bm{k}}\left(v_{\nu, \bm{k}, i}\epsilon_{dlj}m_{\nu, \bm{k}, d}-v_{\nu, \bm{k}, j}\epsilon_{dli}m_{\nu, \bm{k}, d}\right)\frac{df\left(\epsilon_{\nu, \bm{k}}\right)}{d\epsilon_{\nu, \bm{k}}}.
\end{align}
From Eq. \ref{Free_energy_density}, we can then obtain the current density:
\begin{align}
J_i=&-\frac{\partial\Delta F}{\partial A_i}=-\frac{e^2}{\hbar^2}\sum_\nu\int_{\textrm{BZ}}v_{\nu, \bm{k}, i}v_{\nu, \bm{k}, j}\left\{\frac{\tau}{1-i\omega\tau}\frac{df\left(\epsilon_{\nu, \bm{k}}\right)}{d\epsilon_{\nu, \bm{k}}}E_j+\left[\frac{df\left(\epsilon_{\nu, \bm{k}}\right)}{d\epsilon_{\nu, \bm{k}}}-\frac{df\left(\xi_{\nu, \bm{k}}\right)}{d\xi_{\nu, \bm{k}}}\right]A_j\right\},
\end{align}

Interestingly, we notice that the free energy density related to $Q^{\textrm{sta}}_{ijl}$ and $Q^{\textrm{dyn}}_{ijl}$ can be further simplified as
\begin{align}\nonumber
\Delta F_{Q}=&-\frac{i}{V}\frac{e}{\hbar}\sum_{\nu, \bm{k}, \bm{q}, m}\frac{1}{2}A_i\left(-\bm{q}, -i\omega_m\right)\left[\frac{df\left(\xi_{\nu, \bm{k}}\right)}{d\xi_{\nu, \bm{k}}}-\frac{df\left(\epsilon_{\nu, \bm{k}}\right)}{d\epsilon_{\nu, \bm{k}}}\right]\left(v_{\nu, \bm{k}, i}m_{\nu, \bm{k}, d}\epsilon_{dlj}-v_{\nu, \bm{k}, j}m_{\nu, \bm{k}, d}\epsilon_{dli}\right)q_lA_j\left(\bm{q}, i\omega_m\right)\\\nonumber
=&-\frac{1}{2}\sum_{\nu, \bm{q}, \omega}B_i\left(-\bm{q}, -\omega\right)\frac{e}{\hbar}\int_{\textrm{BZ}}\frac{d\bm{k}}{\left(2\pi\right)^d}m_{\nu, \bm{k}, i}v_{\nu, \bm{k}, j}\left[\frac{df\left(\xi_{\nu, \bm{k}}\right)}{d\xi_{\nu, \bm{k}}}-\frac{df\left(\epsilon_{\nu, \bm{k}}\right)}{d\epsilon_{\nu, \bm{k}}}\right]A_j\left(\bm{q}, \omega\right)\\\nonumber
&+\frac{1}{2}\sum_{\nu, \bm{q}, \omega}B_i\left(-\bm{q}, -\omega\right)\frac{e}{\hbar}\frac{i\omega\tau}{1-i\omega\tau}\int_{\textrm{BZ}}\frac{d\bm{k}}{\left(2\pi\right)^d}m_{\nu, \bm{k}, i}v_{\nu, \bm{k}, j}\frac{df\left(\epsilon_{\nu, \bm{k}}\right)}{d\epsilon_{\nu, \bm{k}}}A_j\left(\bm{q}, \omega\right)\\\nonumber
&-\frac{1}{2}\sum_{\nu, \bm{q}, \omega}A_i\left(-\bm{q}, \omega\right)\frac{e}{\hbar}\int_{\textrm{BZ}}\frac{d\bm{k}}{\left(2\pi\right)^d}v_{\nu, \bm{k}, i}m_{\nu, \bm{k}, j}\left[\frac{df\left(\xi_{\nu, \bm{k}}\right)}{d\xi_{\nu, \bm{k}}}-\frac{df\left(\epsilon_{\nu, \bm{k}}\right)}{d\epsilon_{\nu, \bm{k}}}\right]B_j\left(\bm{q}, \omega\right)\\
&+\frac{1}{2}\sum_{\nu, \bm{q}, \omega}A_i\left(-\bm{q}, -\omega\right)\frac{e}{\hbar}\frac{i\omega\tau}{1-i\omega\tau}\int_{\textrm{BZ}}\frac{d\bm{k}}{\left(2\pi\right)^d}v_{\nu, \bm{k}, i}m_{\nu, \bm{k}, j}\frac{df\left(\epsilon_{\nu, \bm{k}}\right)}{d\epsilon_{\nu, \bm{k}}}B_j\left(\bm{q}, \omega\right).
\end{align}
Then, we derive the current-induced orbital magnetization:
\begin{align}
m_i=-\frac{\partial\Delta F}{\partial B_i}=-\frac{e}{\hbar}\sum_\nu\int_{\textrm{BZ}}m_{\nu, \bm{k}, i}v_{\nu, \bm{k}, j}\left\{\frac{\tau}{1-i\omega\tau}\frac{df\left(\epsilon_{\nu, \bm{k}}\right)}{d\epsilon_{\nu, \bm{k}}}E_j+\left[\frac{df\left(\epsilon_{\nu, \bm{k}}\right)}{d\epsilon_{\nu, \bm{k}}}-\frac{df\left(\xi_{\nu, \bm{k}}\right)}{d\xi_{\nu, \bm{k}}}\right]A_j\right\}.
\end{align}
Here the electric field has been expressed in terms of the vector gauge potential as $\bm{E}=i\omega\bm{A}$.

\subsection{Spin Magnetoelectric Susceptibility Calculation}
The spin magnetoelectric susceptibility $T^{\textrm{s}}_{ij}$ is calculated as:
\begin{align}
T^{\textrm{s}}_{ij}=&-\frac{1}{\beta V}\sum_{\bm{k}, n, \nu, \nu'}\frac{\left(\omega_n^2+\omega_n\omega_m-\xi_{\nu, \bm{k}}\xi_{\nu', \bm{k}}-\Delta_{\nu, \bm{k}}\Delta^\ast_{\nu', \bm{k}}\right)\bra{\phi_{\nu, \bm{k}}}\frac{1}{2}\mu_{\textrm{B}}\sigma_i\ket{\phi_{\nu', \bm{k}}}\bra{\phi_{\nu', \bm{k}}}\frac{e}{\hbar}\partial_{k_j}H_0\left(\bm{k}\right)\ket{\phi_{\nu, \bm{k}}}}{\left[\omega_n^2+\xi^2_{\nu, \bm{k}}+\Delta^\ast_{\nu, \bm{k}}\Delta_{\nu, \bm{k}}\right]\left[\left(\omega_n+\omega_m\right)^2+\xi^2_{\nu', \bm{k}}+\Delta^\ast_{\nu', \bm{k}}\Delta_{\nu', \bm{k}}\right]}.
\end{align}
We further sum over the Matsubara frequency and decompose the spin magnetoelectric susceptibility into the static $\left(i\omega_m=0\right)$ and dynamic $\left(i\omega_m\neq0\right)$ parts as
\begin{align}
T^{\textrm{s}}_{ij}=&T^{\textrm{s}, \textrm{sta}}_{ij}+T^{\textrm{s}, \textrm{dyn}}_{ij}\left(i\omega_m\right),
\end{align}
with
\begin{align}\nonumber
T^{\textrm{s}, \textrm{sta}}_{ij}=&-\frac{1}{V}\sum_{\bm{k}, \nu, \nu'}\left[\frac{1}{2}\left(1-\frac{\xi_{\nu, \bm{k}}\xi_{\nu', \bm{k}}+\Delta_{\nu, \bm{k}}\Delta^\ast_{\nu', \bm{k}}}{\epsilon_{\nu', \bm{k}}\epsilon_{\nu, \bm{k}}}\right)\frac{1-f\left(\epsilon_{\nu', \bm{k}}\right)-f\left(\epsilon_{\nu, \bm{k}}\right)}{\epsilon_{\nu', \bm{k}}+\epsilon_{\nu, \bm{k}}}-\frac{1}{2}\left(1+\frac{\xi_{\nu, \bm{k}}\xi_{\nu', \bm{k}}+\Delta_{\nu, \bm{k}}\Delta^\ast_{\nu', \bm{k}}}{\epsilon_{\nu', \bm{k}}\epsilon_{\nu, \bm{k}}}\right)\frac{f\left(\epsilon_{\nu', \bm{k}}\right)-f\left(\epsilon_{\nu, \bm{k}}\right)}{\epsilon_{\nu', \bm{k}}-\epsilon_{\nu, \bm{k}}}\right]\\
&\bra{\phi_{\nu, \bm{k}}}\frac{1}{2}\mu_{\textrm{B}}g\sigma_i\ket{\phi_{\nu', \bm{k}}}\bra{\phi_{\nu', \bm{k}}}\frac{e}{\hbar}\partial_{k_j}H_0\left(\bm{k}\right)\ket{\phi_{\nu, \bm{k}}},
\end{align}
and
\begin{align}\nonumber
T^{\textrm{s}, \textrm{dyn}}_{ij}\left(i\omega_m\right)=&-\frac{1}{V}\sum_{\bm{k}, \nu, \nu'}\left\{\frac{1}{4}\left(1-\frac{\xi_{\nu, \bm{k}}\xi_{\nu', \bm{k}}+\Delta_{\nu, \bm{k}}\Delta^\ast_{\nu', \bm{k}}}{\epsilon_{\nu, \bm{k}}\epsilon_{\nu', \bm{k}}}\right)\left[\frac{i\omega_m}{i\omega_m+\epsilon_{\nu', \bm{k}}+\epsilon_{\nu, \bm{k}}}+\frac{i\omega_m}{i\omega_m-\left(\epsilon_{\nu', \bm{k}}+\epsilon_{\nu, \bm{k}}\right)}\right]\frac{1-f\left(\epsilon_{\nu', \bm{k}}\right)-f\left(\epsilon_{\nu, \bm{k}}\right)}{\epsilon_{\nu', \bm{k}}+\epsilon_{\nu, \bm{k}}}\right.\\\nonumber
&\left.\frac{1}{4}\left(1+\frac{\xi_{\nu, \bm{k}}\xi_{\nu', \bm{k}}+\Delta_{\nu, \bm{k}}\Delta^\ast_{\nu', \bm{k}}}{\epsilon_{\nu', \bm{k}}\epsilon_{\nu, \bm{k}}}\right)\left[\frac{i\omega_m}{i\omega_m-\left(\epsilon_{\nu', \bm{k}}-\epsilon_{\nu, \bm{k}}\right)}+\frac{i\omega_m}{i\omega_m+\epsilon_{\nu', \bm{k}}-\epsilon_{\nu, \bm{k}}}\right]\frac{f\left(\epsilon_{\nu', \bm{k}}\right)-f\left(\epsilon_{\nu, \bm{k}}\right)}{\epsilon_{\nu', \bm{k}}-\epsilon_{\nu, \bm{k}}}\right\}\\
&\bra{\phi_{\nu, \bm{k}}}\frac{1}{2}\mu_{\textrm{B}}g\sigma_i\ket{\phi_{\nu', \bm{k}}}\bra{\phi_{\nu', \bm{k}}}\frac{e}{\hbar}\partial_{k_j}H_0\left(\bm{k}\right)\ket{\phi_{\nu, \bm{k}}}.
\end{align}

The intra-band contribution to $T^{\textrm{s}, \textrm{sta}}_{ij}$ is
\begin{align}
T^{\textrm{s}, \textrm{sta}, \textrm{intra}}_{ij}=&\frac{1}{V}\sum_{\bm{k}, \nu}\frac{df\left(\epsilon_{\nu, \bm{k}}\right)}{d\epsilon_{\nu, \bm{k}}}\bra{\phi_{\nu, \bm{k}}}\frac{1}{2}\mu_{\textrm{B}}g\sigma_i\ket{\phi_{\nu, \bm{k}}}\bra{\phi_{\nu, \bm{k}}}\frac{e}{\hbar}\partial_{k_i}H_0\left(\bm{k}\right)\ket{\phi_{\nu, \bm{k}}}.
\end{align}
For the inter-band contribution, similarly, we assume that the pairing gap $|\Delta_{\nu, \bm{k}}|$ is much smaller than the band splitting so that it can be approximated by that in the normal metal phase. We also know that the summation of the intra-band and inter-band terms in the static limit will vanish, so the inter-band contribution can be written as
\begin{align}
T^{\textrm{s}, \textrm{sta}, \textrm{inter}}_{ij}=&-\frac{1}{V}\sum_{\bm{k}, \nu}\frac{df\left(\xi_{\nu, \bm{k}}\right)}{d\xi_{\nu, \bm{k}}}\bra{\phi_{\nu, \bm{k}}}\frac{1}{2}\mu_{\textrm{B}}g\sigma_i\ket{\phi_{\nu, \bm{k}}}\bra{\phi_{\nu, \bm{k}}}\frac{e}{\hbar}\partial_{k_j}H_0\left(\bm{k}\right)\ket{\phi_{\nu, \bm{k}}}.
\end{align}
We know that for a Bloch state, the spin magnetic moment is $\bm{S}_{\nu, \bm{k}}=\bra{\phi_{\nu, \bm{k}}}\frac{1}{2}\mu_{\textrm{B}}g\bm{\sigma}\ket{\phi_{\nu, \bm{k}}}$, so the static part of the spin magnetoelectric susceptibility can be calculated as
\begin{align}\nonumber
T^{\textrm{s}, \textrm{sta}}_{ij}=&T^{\textrm{s}, \textrm{sta}, \textrm{intra}}_{ij}+T^{\textrm{s}, \textrm{sta}, \textrm{inter}}_{ij}\\
=&-\frac{e}{\hbar}\int_{\textrm{BZ}}\frac{d\bm{k}}{\left(2\pi\right)^d}\sum_{\nu}S_{\nu, \bm{k}, i}v_{\nu, \bm{k}, j}\left[\frac{df\left(\xi_{\nu, \bm{k}}\right)}{d\xi_{\nu, \bm{k}}}-\frac{df\left(\epsilon_{\nu, \bm{k}}\right)}{d\epsilon_{\nu, \bm{k}}}\right].
\end{align}
$\tilde{T}^{\textrm{s}, \textrm{sta}}_{ij}$ can be obtained similarly following the above procedure:
\begin{align}
\tilde{T}^{\textrm{s}, \textrm{sta}}_{ij}=&-\frac{e}{\hbar}\int_{\textrm{BZ}}\frac{d\bm{k}}{\left(2\pi\right)^d}\sum_{\nu}v_{\nu, \bm{k}, i}S_{\nu, \bm{k}, j}\left[\frac{df\left(\xi_{\nu, \bm{k}}\right)}{d\xi_{\nu, \bm{k}}}-\frac{df\left(\epsilon_{\nu, \bm{k}}\right)}{d\epsilon_{\nu, \bm{k}}}\right].
\end{align}

For the dynamic part that is $i\omega_m$ dependent, we consider the dominant intra-band contribution and obtain
\begin{align}
T^{\textrm{s}, \textrm{dyn}}_{ij}\left(i\omega_m\right)=&\frac{i\omega\tau}{1-i\omega\tau}\frac{e}{\hbar}\int_{\textrm{BZ}}\frac{d\bm{k}}{\left(2\pi\right)^d}\sum_\nu S_{\nu, \bm{k}, i}v_{\nu, \bm{k}, j}\frac{df\left(\epsilon_{\nu, \bm{k}}\right)}{d\epsilon_{\nu, \bm{k}}}.
\end{align}
$\tilde{T}^{\textrm{s}, \textrm{dyn}}_{ij}\left(i\omega_m\right)$ can be obtained similarly following the above procedure as well:
\begin{align}
\tilde{T}^{\textrm{s}, \textrm{dyn}}_{ij}\left(i\omega_m\right)=&\frac{i\omega\tau}{1-i\omega\tau}\frac{e}{\hbar}\int_{\textrm{BZ}}\frac{d\bm{k}}{\left(2\pi\right)^d}\sum_\nu v_{\nu, \bm{k}, i}S_{\nu, \bm{k}, j}\frac{df\left(\epsilon_{\nu, \bm{k}}\right)}{d\epsilon_{\nu, \bm{k}}}.
\end{align}

Eventually, we get the spin magnetoelectric susceptibility as
\begin{align}
T_{ij}^{\textrm{s}}=&-\frac{e}{\hbar}\int_{\textrm{BZ}}\frac{d\bm{k}}{\left(2\pi\right)^d}\sum_\nu S_{\nu, \bm{k}, i}v_{\nu, \bm{k}, j}\left[\frac{df\left(\xi_{\nu, \bm{k}}\right)}{d\xi_{\nu, \bm{k}}}-\frac{df\left(\epsilon_{\nu, \bm{k}}\right)}{d\epsilon_{\nu, \bm{k}}}\right]+\frac{i\omega\tau}{1-i\omega\tau}\frac{e}{\hbar}\int_{\textrm{BZ}}\frac{d\bm{k}}{\left(2\pi\right)^d}\sum_\nu S_{\nu, \bm{k}, i}v_{\nu, \bm{k}, j}\frac{df\left(\epsilon_{\nu, \bm{k}}\right)}{d\epsilon_{\nu, \bm{k}}},\\
\tilde{T}_{ij}^{\textrm{s}}=&-\frac{e}{\hbar}\int_{\textrm{BZ}}\frac{d\bm{k}}{\left(2\pi\right)^d}\sum_\nu v_{\nu, \bm{k}, i}S_{\nu, \bm{k}, j}\left[\frac{df\left(\xi_{\nu, \bm{k}}\right)}{d\xi_{\nu, \bm{k}}}-\frac{df\left(\epsilon_{\nu, \bm{k}}\right)}{d\epsilon_{\nu, \bm{k}}}\right]+\frac{i\omega\tau}{1-i\omega\tau}\frac{e}{\hbar}\int_{\textrm{BZ}}\frac{d\bm{k}}{\left(2\pi\right)^d}\sum_\nu v_{\nu, \bm{k}, i}S_{\nu, \bm{k}, j}\frac{df\left(\epsilon_{\nu, \bm{k}}\right)}{d\epsilon_{\nu, \bm{k}}}.
\end{align}

\subsection{Current-induced Total Magnetization}
Now we are ready to write down the expression for the total magnetization induced by current:
\begin{align}
M_i=&-\frac{\partial\Delta F}{\partial B_i}=-\frac{e}{\hbar}\int_{\textrm{BZ}}\frac{d\bm{k}}{\left(2\pi\right)^d}\sum_\nu M_{\nu, \bm{k}, i}v_{\nu, \bm{k}, j}\left\{\frac{\tau}{1-i\omega\tau}\frac{df\left(\epsilon_{\nu, \bm{k}}\right)}{d\epsilon_{\nu, \bm{k}}}E_j+\left[\frac{df\left(\epsilon_{\nu, \bm{k}}\right)}{d\epsilon_{\nu, \bm{k}}}-\frac{df\left(\xi_{\nu, \bm{k}}\right)}{d\xi_{\nu, \bm{k}}}\right]A_j\right\},
\end{align}
which is induced by the current density
\begin{align}
J_i=&-\frac{\partial\Delta F}{\partial A_i}=-\frac{e^2}{\hbar^2}\int_{\textrm{BZ}}\frac{d\bm{k}}{\left(2\pi\right)^d}\sum_\nu v_{\nu, \bm{k}, i}v_{\nu, \bm{k}, j}\left\{\frac{\tau}{1-i\omega\tau}\frac{df\left(\epsilon_{\nu, \bm{k}}\right)}{d\epsilon_{\nu, \bm{k}}}E_j+\left[\frac{df\left(\epsilon_{\nu, \bm{k}}\right)}{d\epsilon_{\nu, \bm{k}}}-\frac{df\left(\xi_{\nu, \bm{k}}\right)}{d\xi_{\nu, \bm{k}}}\right]A_j\right\}.
\end{align}
The total magnetic moment $\bm{M}_{\nu, \bm{k}}=\bm{m}_{\nu, \bm{k}}+\bm{S}_{\nu, \bm{k}}$ includes both the orbital magnetic moment $\bm{m}_{\nu, \bm{k}}$ and the spin magnetic moment $\bm{S}_{\nu, \bm{k}}$. 

As a result, at given current density $\bm{J}$, the induced magnetization becomes $M_i=\alpha_{ik}J_k$, where the susceptibility $\alpha_{ik}$ is
\begin{align}
\alpha_{ik}=\gamma_{ij}\left(\tilde{v}^{-1}\right)_{jk},
\end{align}
with
\begin{align}
\gamma_{ij}=-\frac{e}{\hbar}\sum_\nu\int_{\textrm{BZ}}M_{\nu, \bm{k}, i}v_{\nu, \bm{k}, j}\left[\frac{1}{1-i\omega\tau}\frac{df\left(\epsilon_{\nu, \bm{k}}\right)}{d\epsilon_{\nu, \bm{{k}}}}-\frac{df\left(\xi_{\nu, \bm{k}}\right)}{d\xi_{\nu, \bm{k}}}\right]\frac{d\bm{k}}{\left(2\pi\right)^d},
\end{align}
and
\begin{align}
\tilde{v}_{jk}=-\frac{e^2}{\hbar^2}\sum_\nu\int_{\textrm{BZ}}v_{\nu, \bm{k}, i}v_{\nu, \bm{k}, j}\left[\frac{1}{1-i\omega\tau}\frac{df\left(\epsilon_{\nu, \bm{k}}\right)}{d\epsilon_{\nu, \bm{k}}}-\frac{df\left(\xi_{\nu, \bm{k}}\right)}{d\xi_{\nu, \bm{k}}}\right]\frac{d\bm{k}}{\left(2\pi\right)^d}.
\end{align}
At the zero temperature $T=0$, we know $\frac{df\left(\epsilon_{\nu, \bm{k}}\right)}{d\epsilon_{\nu, \bm{k}}}=0$, so the two tensors $\gamma_{ij}$, $\tilde{v}_{ij}$ become
\begin{align}
\gamma_{ij}=&\frac{e}{\hbar}\sum_\nu\int_{\textrm{BZ}}M_{\nu, \bm{k}, i}v_{\nu, \bm{k}, j}\frac{df\left(\xi_{\nu, \bm{k}}\right)}{d\xi_{\nu, \bm{k}}}\frac{d\bm{k}}{\left(2\pi\right)^d}\sim\frac{e}{\hbar}\sum_\nu\oint M_{\nu, \bm{k}_{\textrm{F}}, i}v_{\nu, \bm{k}_{\textrm{F}}, j}d\bm{k}_{\textrm{F}},
\end{align}
and
\begin{align}
\tilde{v}_{ij}=&\frac{e^2}{\hbar^2}\sum_\nu\int_{\textrm{BZ}}v_{\nu, \bm{k}, i}v_{\nu, \bm{k}, j}\frac{df\left(\xi_{\nu, \bm{k}}\right)}{d\xi_{\nu, \bm{k}}}\frac{d\bm{k}}{\left(2\pi\right)^d}\sim\frac{e^2}{\hbar^2}\sum_\nu\oint v_{\nu, \bm{k}_{\textrm{F}}, i}v_{\nu, \bm{k}_{\textrm{F}}, j}d\bm{k}_{\textrm{F}}.
\end{align}
Near the critical temperature $T_{\textrm{c}}$, we consider the limit $\omega\tau\ll1$, so the tensor $\gamma_{ij}$, $\tilde{v}_{ij}$ can have the taylor expansion in terms of the pairing $\Delta_{\nu, \bm{k}}$ as
\begin{align}\nonumber
\gamma_{ij}\approx&\frac{e}{\hbar}\sum_\nu\int_{\textrm{BZ}}M_{\nu, \bm{k}, i}v_{\nu,  \bm{k}, j}\left[\frac{df\left(\epsilon_{\nu, \bm{k}}\right)}{d\epsilon_{\nu, \bm{k}}}-\frac{df\left(\xi_{\nu, \bm{k}}\right)}{d\xi_{\nu, \bm{k}}}\right]\frac{d\bm{k}}{\left(2\pi\right)^d}\\\nonumber
\approx&\frac{e}{\hbar}\sum_\nu\int_{\textrm{BZ}}M_{\nu, \bm{k}, i}v_{\nu, \bm{k}, j}\Delta^2_{\nu, \bm{k}}\frac{d^2f\left(\epsilon_{\nu, \bm{k}}\right)}{d\epsilon^2_{\nu, \bm{k}}}\frac{d\epsilon_{\nu, \bm{k}}}{d\left(\Delta^2_{\nu, \bm{k}}\right)}|_{\Delta_{\nu, \bm{k}}=0}\\
\sim&\frac{e}{\hbar}\sum_\nu\oint M_{\nu, \bm{k}_{\textrm{F}}, i}v_{\nu, \bm{k}_{\textrm{F}}, j}\Delta^2_{\nu, \bm{k}_{\textrm{F}}}d\bm{k}_{\textrm{F}},
\end{align}
and
\begin{align}\nonumber
\tilde{v}_{ij}\approx&\frac{e^2}{\hbar^2}\sum_\nu\int_{\textrm{BZ}}v_{\nu, \bm{k}, i}v_{\nu, \bm{k}, j}\left[\frac{df\left(\epsilon_{\nu, \bm{k}}\right)}{d\epsilon_{\nu, \bm{k}}}-\frac{df\left(\xi_{\nu, \bm{k}}\right)}{d\xi_{\nu, \bm{k}}}\right]\frac{d\bm{k}}{\left(2\pi\right)^d}\\\nonumber
\approx&\frac{e^2}{\hbar^2}\sum_\nu\int_{\textrm{BZ}}v_{\nu, \bm{k}, i}v_{\nu, \bm{k}, j}\Delta^2_{\nu, \bm{k}}\frac{d^2f\left(\epsilon_{\nu, \bm{k}}\right)}{d\epsilon^2_{\nu, \bm{k}}}\frac{d\epsilon_{\nu, \bm{k}}}{d\left(\Delta^2_{\nu, \bm{k}}\right)}|_{\Delta_{\nu, \bm{k}}=0}\\
\sim&\frac{e^2}{\hbar^2}\sum_\nu\oint v_{\nu, \bm{k}_{\textrm{F}}, i}v_{\nu, \bm{k}_{\textrm{F}}, j}\Delta^2_{\nu, \bm{k}_{\textrm{F}}}d\bm{k}_{\textrm{F}},
\end{align}
with $\bm{k}_{\textrm{F}}$ being the wave vector on the Fermi surfaces. At $T>T_{\textrm{c}}$, the pairing gap $\Delta_{\nu, \bm{k}}$ is zero, so the tensors $\gamma_{ij}$,  $\tilde{v}_{ij}$ are
\begin{align}
\gamma_{ij}=&\frac{e}{\hbar}\sum_{\nu}\int_{\textrm{BZ}}M_{\nu, \bm{k}, i}v_{\nu, \bm{k}, j}\frac{df\left(\xi_{\nu, \bm{k}}\right)}{d\xi_{\nu, \bm{k}}}\frac{d\bm{k}}{\left(2\pi\right)^d}\sim\frac{e}{\hbar}\sum_\nu\oint M_{\nu, \bm{k}_{\textrm{F}}, i}v_{\nu, \bm{k}_{\textrm{F}}, j}d\bm{k}_{\textrm{F}},
\end{align}
and
\begin{align}
\tilde{v}_{ij}=&\frac{e^2}{\hbar^2}\sum_\nu\int_{\textrm{BZ}}v_{\nu, \bm{k}, i}v_{\nu, \bm{k}, j}\frac{df\left(\xi_{\nu, \bm{k}}\right)}{d\xi_{\nu, \bm{k}}}\frac{d\bm{k}}{\left(2\pi\right)^d}\sim\frac{e^2}{\hbar^2}\sum_\nu\oint v_{\nu, \bm{k}_{\textrm{F}}, i}v_{\nu, \bm{k}_{\textrm{F}}, j}d\bm{k}_{\textrm{F}}.
\end{align}

\section{Continuum Model for the Twisted Bilayer Graphene}
In the monolayer graphene, the primitive lattice vectors and the corresponding reciprocal primitive lattice vectors are
\begin{align}
\bm{a}^0_1=&\sqrt{3}\left(\frac{1}{2}, \frac{\sqrt{3}}{2}\right)d,\quad\quad\quad\bm{a}^0_2=\sqrt{3}\left(-\frac{1}{2}, \frac{\sqrt{3}}{2}\right)d,\\
\bm{b}^0_1=&\frac{4\pi}{3d}\left(\frac{\sqrt{3}}{2}, \frac{1}{2}\right),\quad\quad\quad\bm{b}^0_2=\frac{4\pi}{3d}\left(-\frac{\sqrt{3}}{2}, \frac{1}{2}\right),
\end{align}
with $d=1.42\AA$. We also define the vector that links the origin of the unit cell to its respective sublattice $\alpha=A, B$ atom to be $\bm{\delta}_A=\bm{0}$, $\bm{\delta}_B=d\left(0, 1\right)$. The Dirac points localize at the Brillouin zone corner $\bm{K}_\pm=\frac{4\pi}{3d}\left(\frac{\sqrt{3}}{2}, \frac{1}{2}\right)$. Given the uniaxial strain tensor $\bm{\mathcal{E}}$ along the zig-zag direction of one graphene layer, where
\begin{align}
\bm{\mathcal{E}}=\varepsilon\begin{pmatrix}
-1 & 0 \\
0 & \nu_{\textrm{poi}}
\end{pmatrix},
\end{align}
with the Poison's ratio $\nu_{\textrm{poi}}=0.165$, we know that the uniaxial strain will deform both the real and reciprocal space as
\begin{align}
\tilde{\bm{r}}=\left(1+\bm{\mathcal{E}}\right)\bm{r},\quad\quad\quad\tilde{\bm{k}}=\left(1-\bm{\mathcal{E}}^{\textrm{T}}\right)\bm{k}_0.
\end{align}
As a result, the strain changes the position of the Dirac points in the reciprocal space to be
\begin{align}
\tilde{\bm{K}}_{\eta}=\left(1-\bm{\mathcal{E}}^{\textrm{T}}\right)\bm{K}_{\eta}-\eta\bm{A}_{\textrm{strain}},
\end{align}
with the valley index $\eta=\pm1$ and the strain induced effective gauge field $\bm{A}_{\textrm{strain}}=\frac{\beta}{d}\left(\mathcal{E}_{xx}-\mathcal{E}_{yy}, -2\mathcal{E}_{xy}\right)$, $\beta=1.57$. The Hamiltonian for the bottom layer graphene at the valley $\eta$ then becomes
\begin{align}
\tilde{\mathcal{H}}_{\textrm{b}}=\sum_{\bm{k}, s, \eta}c^\dagger_{\textrm{b}, s, \eta}\left(\bm{k}\right)h_{\textrm{b}, \eta}\left(\bm{k}\right)c_{\textrm{b}, s, \eta}\left(\bm{k}\right)=\sum_{\bm{k}, s, \eta}c^\dagger_{\textrm{b}, s, \eta}\left[\eta\hbar v_{\textrm{F}}\hat{\bm{R}}_{-\frac{\theta}{2}}\left(1+\bm{\mathcal{E}}^{\textrm{T}}\right)\left(\bm{k}+\eta\bm{A}_{\textrm{strain}}\right)\cdot\bm{\sigma}+\Delta\sigma_z\right]c_{\textrm{b}, s, \eta}\left(\bm{k}\right),
\end{align}
with the spin index $s=\uparrow, \downarrow$, the rotation matrix $\hat{\bm{R}}_{-\frac{\theta}{2}}=\cos\frac{\theta}{2}+i\sigma_y\sin\frac{\theta}{2}$, and the momentum $\bm{k}=\bm{k}-\left(1-\bm{\mathcal{E}}^{\textrm{T}}\right)\bm{K}_\eta$. The top layer graphene has the Hamiltonian
\begin{align}
\mathcal{H}_{\textrm{t}}=\sum_{\bm{k}, s, \eta}c^\dagger_{\textrm{t}, s, \eta}\left(\bm{k}\right)h_{\textrm{t}, \eta}\left(\bm{k}\right)c_{\textrm{t}, s, \eta}=\sum_{\bm{k}, s, \eta}c^\dagger_{\textrm{t}, s, \eta}\left(\bm{k}\right)\eta\hbar v_{\textrm{F}}\hat{\bm{R}}_{\frac{\theta}{2}}\bm{k}\cdot\bm{\sigma}c_{\textrm{t}, s, \eta}\left(\bm{k}\right).
\end{align}
Then, we consider the tunneling matrix element from the bottom layer to the top layer~\cite{Neto1, Neto2, MacDonald} to be
\begin{align}\nonumber
\tilde{T}^{\alpha, \beta}_{\hat{\bm{R}}_{-\frac{\theta}{2}}\tilde{\bm{K}}_\eta+\bm{k}, \hat{\bm{R}}_{\frac{\theta}{2}}\bm{K}_\eta+\bm{k}'}=&\frac{1}{3}t_\perp\left[\delta_{\hat{\bm{R}}_{-\frac{\theta}{2}}\tilde{\bm{K}}_\eta+\bm{k}, \hat{\bm{R}}_{\frac{\theta}{2}}\bm{K}_\eta+\bm{k}'}+e^{i\tilde{\bm{b}}_2\cdot\left(\tilde{\bm{\delta}}_{\alpha}-\tilde{\bm{\delta}}_{\beta}\right)}\delta_{\hat{R}_{-\frac{\theta}{2}}\left(\tilde{\bm{K}}_\eta+\tilde{\bm{b}}_2\right)+\bm{k}, \hat{\bm{R}}_{\frac{\theta}{2}}\left(\bm{K}_\eta+\bm{b}_2+\right)+\bm{k}'}\right.\\
&\left.+e^{-i\tilde{b}_1\cdot\left(\tilde{\bm{\delta}}_\alpha-\tilde{\bm{\delta}}_\beta\right)}\delta_{\hat{R}_{-\frac{\theta}{2}}\left(\tilde{\bm{K}}_\eta-\tilde{\bm{b}}_1\right)+\bm{k}, \hat{\bm{R}}_{\frac{\theta}{2}}\left(\bm{K}_\eta-\bm{b}_1\right)+\bm{k}'}\right],
\end{align}
so the strain deformed interlayer Hamiltonian becomes
\begin{align}
\tilde{\mathcal{H}}_{\textrm{int}}=&\sum_{\bm{k}, s, \eta}c^\dagger_{\textrm{b}, s, \eta}\left(\bm{k}\right)\left[\tilde{T}_{\eta\tilde{\bm{k}}_{\textrm{b}}}\delta_{\bm{k}'-\bm{k}, \eta\tilde{\bm{k}}_{\textrm{b}}}+\tilde{T}_{\eta\tilde{\bm{k}}_{\textrm{tr}}}\delta_{\bm{k}'-\bm{k}, \eta\tilde{\bm{k}}_{\textrm{tr}}}+\tilde{T}_{\eta\tilde{\bm{k}}_{\textrm{tl}}}\delta_{\bm{k}'-\bm{k}, \eta\tilde{\bm{k}}_{\textrm{tl}}}\right]c_{\textrm{t}, s, \eta}\left(\bm{k}'\right)+h. c.,
\end{align}
where
\begin{align}
\tilde{T}_{\eta\tilde{\bm{k}}_{\textrm{b}}}=&\frac{1}{3}t_\perp\begin{pmatrix}
1 & 1 \\
1 & 1
\end{pmatrix},\\
\tilde{T}_{\eta\tilde{\bm{k}}_{\textrm{tr}}}=&\frac{1}{3}t_\perp\begin{pmatrix}
1 & e^{-i\eta\frac{2\pi}{3}\left(1+\sqrt{3}\mathcal{E}_{xx}\mathcal{E}_{xy}+\sqrt{3}\mathcal{E}_{xy}\mathcal{E}_{yy}-\mathcal{E}^2_{xy}-\mathcal{E}^2_{yy}\right)} \\
e^{i\eta\frac{2\pi}{3}\left(1+\sqrt{3}\mathcal{E}_{xx}\mathcal{E}_{xy}+\sqrt{3}\mathcal{E}_{xy}\mathcal{E}_{yy}-\mathcal{E}^2_{xy}-\mathcal{E}^2_{yy}\right)} & 1
\end{pmatrix},\\
\tilde{T}_{\eta\tilde{\bm{k}}_{\textrm{tl}}}=&\frac{1}{3}t_\perp\begin{pmatrix}
1 & e^{i\eta\frac{2\pi}{3}\left(1-\sqrt{3}\mathcal{E}_{xx}\mathcal{E}_{xy}-\sqrt{3}\mathcal{E}_{xy}\mathcal{E}_{yy}-\mathcal{E}^2_{xy}-\mathcal{E}^2_{yy}\right)} \\
e^{-i\eta\frac{2\pi}{3}\left(1-\sqrt{3}\mathcal{E}_{xx}\mathcal{E}_{xy}-\sqrt{3}\mathcal{E}_{xy}\mathcal{E}_{yy}-\mathcal{E}^2_{xy}-\mathcal{E}^2_{yy}\right)} & 1
\end{pmatrix},
\end{align}
with
\begin{align}\nonumber
\tilde{\bm{k}}_{\textrm{b}}=&\hat{\bm{R}}_{-\frac{\theta}{2}}\left(1-\bm{\mathcal{E}}^{\textrm{T}}\right)\bm{K}_+-\hat{\bm{R}}_{\frac{\theta}{2}}\bm{K}_+\\
=&-\frac{4\pi}{3\sqrt{3}d}\begin{pmatrix}
\mathcal{E}_{xx}\cos\frac{\theta}{2}+\mathcal{E}_{xy}\sin\frac{\theta}{2}, & \left(2-\mathcal{E}_{xx}\right)\sin\frac{\theta}{2}+\mathcal{E}_{xy}\cos\frac{\theta}{2}
\end{pmatrix}\\\nonumber
\tilde{\bm{k}}_{\textrm{tr}}=&\hat{\bm{R}}_{-\frac{\theta}{2}}\left(1-\bm{\mathcal{E}}^{\textrm{T}}\right)\left(\bm{K}_++\bm{b}_2\right)-\hat{\bm{R}}_{\frac{\theta}{2}}\left(\bm{K}_++\bm{b}_2\right)\\
=&\frac{2\pi}{9d}\begin{pmatrix}
\left(\sqrt{3}\mathcal{E}_{xx}-3\mathcal{E}_{xy}\right)\cos\frac{\theta}{2}+\left(6+\sqrt{3}\mathcal{E}_{xy}-3\mathcal{E}_{yy}\right)\sin\frac{\theta}{2} ,& -\left(3\mathcal{E}_{yy}-\sqrt{3}\mathcal{E}_{xy}\right)\cos\frac{\theta}{2}+\left(2\sqrt{3}+3\mathcal{E}_{xy}-\sqrt{3}\mathcal{E}_{xx}\right)\sin\frac{\theta}{2}
\end{pmatrix}\\\nonumber
\tilde{\bm{k}}_{\textrm{tl}}=&\hat{\bm{R}}_{-\frac{\theta}{2}}\left(1-\bm{\mathcal{E}}^{\textrm{T}}\right)\left(\bm{K}_+-\bm{b}_1\right)-\hat{\bm{R}}_{\frac{\theta}{2}}\left(\bm{K}_+-\bm{b}_1\right)\\
=&\frac{2\pi}{9d}\begin{pmatrix}
\left(\sqrt{3}\mathcal{E}_{xx}+3\mathcal{E}_{xy}\right)\cos\frac{\theta}{2}-\left(6-\sqrt{3}\mathcal{E}_{xy}-3\mathcal{E}_{xx}\right)\sin\frac{\theta}{2} ,& \left(3\mathcal{E}_{yy}+\sqrt{3}\mathcal{E}_{xy}\right)\cos\frac{\theta}{2}+\left(2\sqrt{3}-3\mathcal{E}_{xy}-\sqrt{3}\mathcal{E}_{xx}\right)\sin\frac{\theta}{2}
\end{pmatrix}.
\end{align}
The interlayer hopping strength is taken as $t_\perp=0.33$eV. As the uniaxial strain deforms the Moir\'e superlattice, the reciprocal primitive lattice vector for the strained Moir\'e superlattice can be obtained as
\begin{align}\nonumber
\tilde{\bm{b}}_1^m=&\tilde{\bm{k}}_{\textrm{b}}-\tilde{\bm{k}}_{\textrm{tl}}\\
=&\frac{2\pi}{3d}\left(-\left(\sqrt{3}\mathcal{E}_{xx}+\mathcal{E}_{xy}\right)\cos\frac{\theta}{2}+\left(2-
\sqrt{3}\mathcal{E}_{xy}-\mathcal{E}_{yy}\right)\sin\frac{\theta}{2}, -\left(\mathcal{E}_{yy}+\sqrt{3}\mathcal{E}_{xy}\right)\cos\frac{\theta}{2}-\left(2\sqrt{3}-\mathcal{E}_{xy}-\sqrt{3}\mathcal{E}_{xx}\right)\sin\frac{\theta}{2}\right),\\
\tilde{\bm{b}_2}^m=&\tilde{\bm{k}}_{\textrm{tr}}-\tilde{\bm{k}}_b\\
=&\frac{2\pi}{3d}\left(\left(\sqrt{3}\mathcal{E}_{xx}-\mathcal{E}_{xy}\right)\cos\frac{\theta}{2}+\left(2+\sqrt{3}\mathcal{E}_{xy}-\mathcal{E}_{yy}\right)\sin\frac{\theta}{2}, -\left(\mathcal{E}_{yy}-\sqrt{3}\mathcal{E}_{xy}\right)\cos\frac{\theta}{2}+\left(2\sqrt{3}+\mathcal{E}_{xy}-\sqrt{3}\mathcal{E}_{xx}\right)\sin\frac{\theta}{2}\right),
\end{align}
and the primitive lattice vectors $\tilde{\bm{a}}_{1}=\left(\tilde{a}_{1x}, \tilde{a}_{1y}\right)$, $\tilde{\bm{a}}_2=\left(\tilde{a}_{2x}, \tilde{a}_{2y}\right)$ are obtained by
\begin{align}
\begin{pmatrix}
\tilde{a}_{1x} & \tilde{a}_{2x} \\
\tilde{a}_{1y} & \tilde{a}_{2y}
\end{pmatrix}=\begin{pmatrix}
\tilde{b}^m_{1x} & \tilde{b}^m_{1y} \\
\tilde{b}^m_{2x} & \tilde{b}^m_{2y}
\end{pmatrix}^{-1}\begin{pmatrix}
2\pi & 0 \\
0 & 2\pi
\end{pmatrix}.
\end{align}
Finally, the Hamiltonian for the twisted bilayer graphene aligned with boron nitride substrate is written as
\begin{align}\nonumber
\mathcal{H}=&\tilde{\mathcal{H}_{\textrm{b}}}+\mathcal{H}_{\textrm{t}}+\tilde{\mathcal{H}}_{\textrm{int}}\\
=&\sum_{\bm{k}, s, \eta}A^\dagger_{s, \eta}\left(\bm{k}\right)h_{\eta}\left(\bm{k}\right)A_{s, \eta}\left(\bm{k}\right),
\end{align}
where $A_{s, \eta}\left(\bm{k}\right)$ has infinite components representing the series of states $c_{\textrm{b}, s, \eta}\left(\bm{k}\right)$, $c_{\textrm{t}, s, \eta}\left(\bm{k}'\right)$ with $\bm{k}-\bm{k}'=\eta\bm{k}_{\textrm{b}}, \eta\bm{k}_{\textrm{tr}}, \eta\bm{k}_{\textrm{tl}}$. The Hamiltonian matrix $h_\eta\left(\bm{k}\right)$ in the truncated basis $\left[a_{\textrm{b}, s, \eta}\left(\bm{k}\right), a_{\textrm{t}, s, \eta}\left(\bm{k}+\eta\bm{k}_{\textrm{b}}\right), a_{\textrm{t}, s, \eta}\left(\bm{k}+\eta\bm{k}_{\textrm{tr}}\right), a_{\textrm{t}, s, \eta}\left(\bm{k}+\eta\bm{k}_{tl}\right)\right]^{\textrm{T}}$ then has the form 
\begin{align}
h_\eta\left(\bm{k}\right)=\begin{pmatrix}
h_{\textrm{b}, \eta}\left(\bm{k}\right) & \tilde{T}_{\eta\tilde{\bm{k}}_{\textrm{b}}} & \tilde{T}_{\eta\tilde{\bm{k}}_{\textrm{tr}}} & \tilde{T}_{\eta\tilde{\bm{k}}_{\textrm{tl}}}\\ 
 \tilde{T}_{\eta\tilde{\bm{k}}_{\textrm{b}}}^\dagger & h_{\textrm{t}, \eta}\left(\bm{k}+\eta\tilde{\bm{k}}_{\textrm{b}}\right) & 0 & 0\\ 
\tilde{T}_{\eta\tilde{\bm{k}}_{\textrm{tr}}}^\dagger & 0 & h_{\textrm{t}, \eta}\left(\bm{k}+\eta\tilde{\bm{k}}_{\textrm{tr}}\right) & 0\\ 
\tilde{T}_{\eta\tilde{\bm{k}}_{\textrm{tl}}}^\dagger & 0 & 0 & h_{\textrm{t}, \eta}\left(\bm{k}+\eta\tilde{\bm{k}}_{\textrm{tl}}\right)
\end{pmatrix}.
\end{align}
We consider 42 sites in the hexagonal reciprocal lattice so that $h_{\eta}\left(\bm{k}\right)$ is an 84$\times$84 matrix in the calculation for the energy dispersion $\xi_{\nu, s, \eta, \bm{k}}$. For a specific band with index $\nu$, the Bogliubov de-Gennes Hamiltonian in the eigen-band basis $\left[\phi^\dagger_{\nu, \uparrow, \eta, \bm{k}}, \phi^\dagger_{\nu, \downarrow, \eta, \bm{k}}, \phi_{\nu, \downarrow, -\eta, -\bm{k}}, \phi_{\nu, \uparrow, -\eta, -\bm{k}}\right]$ can be written as
\begin{align}
\begin{pmatrix}
H_0\left(\bm{k}\right) & \hat{\Delta}\left(\bm{k}\right) \\
\hat{\Delta}^\dagger\left(\bm{k}\right) & -H_0^\ast\left(-\bm{k}\right)
\end{pmatrix}=&\begin{pmatrix}
\xi_{\nu, \uparrow, \eta, \bm{k}} & 0 & \Delta_{\bm{k}} & 0 \\
0 & \xi_{\nu, \downarrow, \eta, \bm{k}} & 0 & -\Delta_{\bm{k}} \\
\Delta_{\bm{k}} & 0 & \xi_{\nu, \downarrow, -\eta, -\bm{k}} & 0 \\
0 & -\Delta_{\bm{k}} & 0 & \xi_{\nu, \uparrow, -\eta, -\bm{k}}
\end{pmatrix},
\end{align}
with the singlet pairing $\Delta_{\bm{k}}=\Delta_0+\lambda\Delta_0\left\{\cos\left(\bm{k}\cdot\tilde{\bm{a}}_1\right)+\cos\left(\bm{k}\cdot\tilde{\bm{a}}_2\right)+\cos\left[\bm{k}\cdot\left(\tilde{\bm{a}}_1-\tilde{\bm{a}}_2\right)\right]\right\}$.

\end{document}